\documentclass{article}

\usepackage{arxiv}

\usepackage[utf8]{inputenc} 
\usepackage[T1]{fontenc}    
\usepackage{hyperref}       
\usepackage{url}            
\usepackage{booktabs}       
\usepackage{amsfonts}       
\usepackage{nicefrac}       
\usepackage{microtype}      
\usepackage{lipsum}
\usepackage{graphicx}
\graphicspath{ {./images/} }

\usepackage{natbib}
\usepackage{placeins}
\usepackage{subfig}
\usepackage{url}
\usepackage{amsmath}
\usepackage{amssymb}
\usepackage{tabularx}
\usepackage{array}
\newcolumntype{Y}{>{\raggedright\arraybackslash}X}
\usepackage{xltabular}
\usepackage{dsfont}  
\usepackage{booktabs}  
\usepackage{graphicx}
\usepackage[font = footnotesize, labelfont = bf]{caption} 
\usepackage{float} 
\usepackage{appendix}

\title{Simulation-based Power Analysis for \\ Sequential Multiple Assignment Randomized Trials}

\author{
 Niki Z. Petrakos \\
  Department of Epidemiology, Biostatistics and Occupational Health \\
  McGill University \\
  Montreal, Canada \\
  \texttt{niki.petrakos@mail.mcgill.ca} \\
   \And
 Erica E. M. Moodie \\
  Department of Epidemiology, Biostatistics and Occupational Health\\
  McGill University\\
  Montreal, Canada \\
  \texttt{erica.moodie@mcgill.ca} \\
  \And
 Nicolas Savy \\
  Institut de Mathématiques de Toulouse; UMR5219 \\
  Université de Toulouse; CNRS. UT2J \\
  Toulouse, France \\
  \texttt{nicolas.savy@math.univ-toulouse.fr} \\
  \And
 Manon de Raad \\
  St. Mary's Research Centre \\
  Montreal, Canada \\
  \texttt{manon.deraad@ssss.gouv.qc.ca} \\
  \And
 Eric Belzile \\
  St. Mary's Research Centre \\
  Montreal, Canada \\
  \texttt{eric.belzile@ssss.gouv.qc.ca} \\
  \And
 Sylvie Lambert \\
  Ingram School of Nursing \\
  McGill University \\
  Montreal, Canada \\
  \texttt{sylvie.lambert@mcgill.ca} \\
}

\begin{document}
\maketitle
\begin{abstract}
Sequential Multiple Assignment Randomized Trials (SMARTs) provide evidence for treatment sequences based on patient profiles, which is relevant in chronic disease settings. Sample size formulae implemented in calculators are the primary tool available to power SMARTs, though they require strong assumptions. We propose a simulation-based procedure omitting these assumptions, instead generating realistic synthetic SMART data by fitting models to real pilot data, to power SMARTs to compare treatment strategies. The proposed framework powers designs in two ways: by fixing the data generating mechanism and estimating effect size under different designs, or by fixing effect size and varying operational decisions within the SMART. Comparing our results to a calculator (SMARTsize), estimated sample sizes at varying power levels were similar at larger fixed effect sizes, whereas a discrepancy was apparent at smaller effect sizes due to differences between fixed and observed effect sizes in the simulated trials. The simulation-based procedure’s ability to capture this effect size fluctuation is advantageous for smaller expected effect sizes, as it is essential to ensure adequate sample size to avoid a type II error. In providing flexible tools to power competing SMART designs, the full potential of SMARTs to build treatment sequences can be better realized.
\end{abstract}


\section{Introduction}
\label{sec:intro}

To test the safety and efficacy of a novel intervention, randomized controlled trials (RCTs) are held as the gold standard, as treatments are randomly assigned to participants, thus allowing for the estimation of the causal effect of a treatment on an outcome \citep{Hariton2018}. Generally, RCTs begin with a baseline visit where participant characteristics are recorded and then proceed to treatment assignment using a pre-determined randomization scheme. Then, participants are monitored for a pre-defined follow-up period, during which intermediate measurements may be recorded, until the trial outcome is measured. Notably, this traditional design involves only one instance (stage) of treatment randomization, which is often sufficient when studying a single, time-invariant treatment. However, there are a multitude of instances, often in chronic disease settings, where a \textit{sequence} of treatments is of interest: switching between antiretroviral therapies to treat human immunodeficiency virus, treatment sequencing in oncology, stepped-care approaches for mental health concerns, among others \citep{Wang2023}. Importantly, this means treatment can vary over time, and thus a standard RCT with a single stage of treatment randomization is no longer sufficient to determine the best treatment sequence for a given patient profile.

A class of trials that allows for the study of pre-defined time-varying treatment sequences is the Sequential Multiple Assignment Randomized Trial, or SMART \citep{Murphy2005,Collins2007,LavoriDawson2014}. The original purpose, and indeed great benefit, of SMARTs is to assess time-varying adaptive interventions considering individual patient characteristics, thus informing personalized medicine. More specifically, SMARTs can be used to compare several \textit{embedded dynamic treatment regimes} (embedded DTRs or EDTRs) or to estimate the optimal EDTR. A DTR is a set of decision rules to adaptively determine which treatment to assign at a given time point given a patient's evolving condition. An EDTR is, by design, a DTR embedded within a SMART where stage 1 treatment response is defined by pre-specified variables (i.e., some intermediate measures, often termed tailoring variables). SMART data can be used to estimate the optimal DTR as well, though this is a more challenging task and thus relevant within the context of exploratory trials; we focus here on confirmatory trials and thus EDTR comparisons. An optimal (E)DTR is one that maximizes the participant's expected outcome given characteristics, treatment assignment, and intermediate outcomes (maximizing the mean outcome over the relevant patient population given the DTR is most common, though optimality can also be defined by minimizing a negative outcome). SMART data can also be used to compare single-stage treatments, such as to determine the effect of the first-stage (initial) treatment \citep{Lambert2023}. A SMART can have more than two stages, but in practice, most existing SMARTs have been limited to two stages since the number of potential treatment trajectories increases rapidly as the number of treatment randomization stages grows, which then limits the observed participants per treatment sequence.

SMARTs provide a wealth of knowledge, as their data can be used to address four aims that target either specific components of the treatment sequence or that consider the entire sequence as a whole \citep{Oetting2007}. These aims include: 1) the treatment effect of stage 1 treatment, 2) the treatment effect of stage 2 treatment to determine the best treatment following completion of stage 1 treatment, 3) the difference in treatment effect between two sequences that are comprised of different first stage treatments, and 4) which treatment strategy yields the best outcome \citep{Lambert2026}. The first two aims are single-stage comparisons (much like a traditional RCT), whereas the latter two aims study entire treatment sequences. These four broad categories of comparisons possible within a SMART are referred to as aims 1-4 for the remainder of the paper. 
 
When designing a trial, it is important to ensure it is well-powered to detect the minimal clinically important difference or MCID (the smallest outcome change that benefits participants), or the standardized effect size (the standardized difference between treatment groups) \citep{ChuangStein2011,Franceschini2023}. For single stage treatment comparisons, simple formulae are often used to calculate the minimum required sample size to detect the MCID or effect size. This is easily done using a sample size calculator, where a user need only input a small number of data features and statistical control parameters: type of outcome (binary, continuous, etc.), type I error probability ($\alpha$), desired level of power ($1-\beta$, where $\beta$ is the type II error probability), and anticipated MCID or effect size. SMARTs also allow for EDTR comparisons, and to power a SMART for this purpose (aim 3), there exist similar sample size formulae and calculators, requiring additional user input for the rate of response to stage 1 treatment \citep{Oetting2007,Kidwell2018}. However, the sample size calculators also require assumptions that are often unrealistic: e.g., treatment groups are of equal size, response rate to first stage treatment is the same across all treatment arms (or that response rate is zero), and the variance of the outcome of a strategy among responders or non-responders is less than the overall variance of the mean outcome of the strategy \citep{Oetting2007}. Additionally, attrition is not factored into these formulae, and hence trialists usually make an educated guess regarding the proportion of drop-out and then inflate the calculated minimum sample size by a suitable factor. This is the current recommendation when powering SMARTs \citep{Petracci2020,Lorenzoni2023}, even though this procedure ignores likely differences in drop-out across treatment stages. Despite these limitations, sample size calculators to power SMARTs to compare EDTRs remain the most user-friendly and accessible option and have been developed for continuous outcomes \citep{Seewald2024}, binary outcomes \citep{Seewald2024,Dziak2024}, clustered SMARTs \citep{NeCamp2017}, and pilot SMARTs \citep{Kim2016}. 

In recent years, simulation tools have become more widely used to power adaptive trials as sample size formulae are unable to account for design considerations that are more data-driven  \citep{Jayawardana2025,Thorlund2018,FDA2019,Mayer2019,Westfall2008}. However, there are very few simulation tools to power full-scale SMARTs to estimate the optimal EDTR (aim 4) \citep{Artman2020}, and to our knowledge, none exist to power a SMART to compare two EDTRs that begin with differing stage 1 treatments (aim 3). In fact, it is not surprising that so few simulation tools exist for powering SMARTs. In practice, SMARTs are almost always powered to perform single-stage comparisons (for which sample size formulae are sufficient), despite their original purpose of and unique ability for learning about treatment sequences. One of the main reasons why it is so difficult to adequately power a SMART to study EDTRs is due to EDTR outcomes being correlated, as different EDTRs can share interventions and participant treatment histories can be consistent with more than one EDTR \citep{Artman2020,Lorenzoni2023}. Hence, there remain significant methodological challenges with regards to powering a design to compare EDTRs \citep{Lorenzoni2023}. \cite{Oetting2007} designed a simulation study to test the robustness of their proposed sample size formulae for powering SMARTs to estimate the optimal EDTR, which assumes a rather simple correlation structure for the mean outcomes per treatment strategy as well as normally-distributed outcomes. More recently, \cite{Artman2020} proposed a Monte Carlo simulation approach to screen out the inferior EDTR of the set compared in the trial by assuming a correlation structure for the EDTR outcomes. When defining the correlation structure, Artman and colleagues recommended using empirical estimates from a pilot SMART, though they note that their power calculations are sensitive to the choice of correlation structure. Most other work focuses on powering small pilot SMARTs, where the aim is to determine feasibility and acceptability and hence is not applicable for confirmatory trials \citep{Almirall2012,Gunlicks-Stoessel2016,Kim2016}. In this work, we focus on filling the gap in the literature of developing a simulation tool to power SMARTs for aim 3.

Most studies are thus powered only for aims 1 or 2 and almost all SMARTs in practice define their primary or secondary aims to be single-stage comparisons \citep{Bigirumurame2022,Lorenzoni2023,Freeman2025}. Even when a SMART defines an aim that involves EDTR comparisons or optimal EDTR estimation, the sample size calculation is almost always based on a single-stage comparison and thus the design is not adequately powered to study EDTRs \citep{Bigirumurame2022}. \cite{Bigirumurame2022} reported in their systematic review that the primary aims of all completed SMARTs included aim 1, whereas only two trials additionally compared EDTRs and one trial estimated the optimal DTR. Of published protocols, the authors found that 21\% (five) included aim 3 and 13\% (three) included aim 4, however nearly all published protocols (96\%) were powered to study only single-stage comparisons \citep{Bigirumurame2022,Lambert2026}. Consequently, SMARTs are still underutilized and the full potential of the SMART design is almost always unrealized \citep{Freeman2025}. Moreover, the number of SMARTs implemented in practice is still limited, despite the great utility of SMARTs to develop evidenced-based, personalized treatment trajectories \citep{Lorenzoni2023}. This is likely due to the aforementioned methodological challenges as well as a lack of established guidelines for designing a SMART \citep{Lorenzoni2023,Freeman2025}. Furthermore, as SMART data can be used to study multiple research questions (e.g., aims 1-4), there is an added complexity, and perhaps confusion in practice, regarding whether a SMART is powered to study all, or only some, of the defined research aims \citep{Freeman2025}. 

Thus far, we have discussed only the difficulties with powering SMARTs. However, other aspects of trial design are also necessary to consider. In a trial with only one stage of treatment randomization, there already exists several design factors to consider \citep{Lambert2026,Lambert2023}. These include participant inclusion and exclusion criteria, treatment comparators (number of arms, dosage, etc.), duration of follow-up (number of weeks, months, years), and definition of the trial outcome. One of the main reasons why trials fail to advance to subsequent testing (e.g., Phase III) or to launching a drug on the market is due to poor design \citep{VanNorman2019}. This is especially problematic given the time and resources needed \citep{Paul2010,VanNorman2016,VanNorman2019}. In a SMART, the added complexity of multiple stages of treatment (re-)randomization introduces even more design considerations, such as duration between treatment stages, definition of responder status (including tailoring variables), and re-randomization schemes (whether all participants or only non-responders are re-randomized). This makes it all the more challenging to define the study design, thus further exacerbating the issue of dedicating enormous time and resources to a trial that may, in the end, return a null result simply due to design inadequacies. 

Some researchers have turned to synthetic data methods to test competing trial designs where trials are run in simulation, also known as ``in silico trials" \citep{Pappalardo2019,Friedrich2024}. In silico trials have been used to determine trial operating characteristics in traditional, single-stage RCTs, such as inclusion and exclusion criteria, sample size, follow-up time, and total study duration \citep{Kim2025}. In the context of SMARTs, one study used historical pilot data to build adaptive interventions to be tested in a subsequent, full-scale trial, where simulated trials differed in intervention, sample size, and study duration \citep{Ferstad2024}. However, the purpose was not to plan a full-scale SMART; rather the aim was to identify a single adaptive intervention to be tested in a standard RCT. Another study used historical pilot data to design a SMART, but competing SMART designs were not tested; the purpose was to determine if a SMART or a traditional RCT was more appropriate to answer the clinical questions of interest \citep{Yan2021}. Rather simplistic generative models were defined to run the simulated trials (e.g., simulating continuous variables by drawing from a univariate Normal with estimated mean and standard deviations from the pilot data), thus diminishing real-world applicability. To our knowledge, pilot SMART data have yet to be employed to determine trial operating characteristics other than sample size for future, full-scale SMARTs, with the aim of comparing EDTRs by using \textit{realistic} data generation methods. 

Thus, there lacks accessible, generalizable methods to power SMARTs and compare competing designs that make it feasible to perform EDTR comparisons while also accounting for attrition. It has been suggested that there is a need for simulation tools that are adaptable to varying SMART designs as the field of trial design is generally moving in this direction \citep{Seewald2020}, and that it would be of great benefit to allow for different response rates to stage 1 treatment, as assuming a single response rate is a strong and unrealistic assumption that can negatively impact SMART implementation \citep{Kim2016}. 

To address this gap, we propose a method to power SMARTs to address aim 3 (comparing two EDTRs with different stage 1 treatments) by utilizing our previously-proposed framework for generating realistic and complex RCT data while also accounting for missing values \citep{Petrakos2025a,Petrakos2025b}. As it is common practice to first run a pilot trial to determine feasibility and acceptability before running a full-scale trial, this simulation-based power analysis is designed to power a full-scale SMART after a pilot has already been completed (and thus pilot data are available). Hence, the proposed procedure requires fitting generative models to pilot SMART data. However, pilots are small, and so we augment pilot SMART baseline data using external data (our real data demonstration uses a historical RCT, but observational data can be utilized as well). The organization of the paper is as follows: Section~\ref{sec:smartdesignnotation} discusses SMART designs and introduces notation, Section~\ref{sec:dataaugmentation} provides a brief overview of external data methods in the trial context and explains how external data are used in our proposed method, Section~\ref{sec:frameworktopowersmart} introduces the real data example, describes the proposed simulation procedure to power SMARTs and compare competing designs, and presents the baseline comparator, Section~\ref{sec:results} gives simulation results and comparisons, and Section~\ref{sec:discussion} concludes with a discussion of strengths, limitations, and potential directions for future work.

\section{SMART Design and Notation}
\label{sec:smartdesignnotation}

A SMART begins with the collection of a set of $\mathbf{X}$ baseline covariates that are relevant to the planned analysis, such as a participant's demographics, medical information, or socioeconomic variables; these data are collected for the $n$ trial participants and generally have very little, if any, missing information. The methods that follow in Section~\ref{sec:frameworktopowersmart} assume no missingness at baseline; we will return to this assumption in Sections~\ref{sec:dataaugmentation} and~\ref{sec:discussion}. Then, participants are randomly assigned to one of the stage 1 treatment arms; generally there are two such arms, which we denote by $A_1=\{0,1\}$. The simplest randomization scheme is having an equal probability of being assigned to either treatment, though stratified or block-randomized designs are also standard. The trial proceeds with the first follow-up period, at the end of which intermediate variables, including tailoring variables, are measured; this time point is denoted $T_1$ and variables collected at $T_1$ are denoted $Z_{1j},j=\{1,...,J\}$ where $J$ is the total number of variables collected at $T_1$. Then, participants are categorized as either responders or non-responders to their $A_1$ assignment, where response status, $R_1$, is a pre-defined, deterministic function of the baseline or tailoring variables. 

At this point, SMARTs generally differ in design in three main ways: either all participants (responders and non-responders) are re-randomized to stage 2 treatment, only the non-responders are re-randomized, or only the responders are re-randomized. If only the responders or non-responders are re-randomized, then those who are not re-randomized either all get assigned to a new treatment or remain on their assigned $A_1$ treatment at stage 2. No matter the choice, some or all participants are then re-randomized to stage 2 treatment, denoted $A_2$. $A_2$ treatment assignments may be entirely new treatments but can also include the same treatment options as $A_1$. Again, the trial proceeds with a follow-up period, and then variables are measured at $T_2$. In a two-stage SMART, this would also be the end of the trial, in which case the outcome $Y$ would also be recorded. A visual flowchart of an example SMART design is included in Section~\ref{subsec:data}.

A strategy, or EDTR, is then one of the sequences of treatment decisions. For example, in a design where only non-responders are re-randomized at stage 2 and responders remain on stage 1 treatment, one strategy is to assign $A_1=1$, followed by $A_2=1$ for responders and $A_2=2$ (some other treatment) for non-responders. Figure~\ref{fig:smartflowchart} shows the four strategies embedded in the example SMART, discussed in Section~\ref{subsec:data}.

\section{Incorporating External Data in Trials}
\label{sec:dataaugmentation}

There has been growing interest in external data, either from historical RCTs or observational data such as electronic health records, in the conduct of clinical trials, with the U.S. Food and Drug Administration releasing guidelines in 2023 for these so-called \textit{externally controlled trials} \citep{FDA2023}. In these trials, external data are used to either supplement (also called \textit{hybrid controlled trials} \citep{Zhu2020}) or replace entirely the control arm of a clinical trial, thus allowing for more trial participants to be assigned to the treatment of interest. This is beneficial since the trial may attain the desired power with fewer enrolled participants, which is particularly relevant in scenarios where trials may be infeasible due to a rare disease setting or resource constraints \citep{Damone2024}. As in a traditional RCT, inference is still the priority in externally or hybrid controlled trials, thus making it of utmost importance that the similarity between internal and external participants is ensured, as external data can introduce bias if they are not consistent with the internal trial data \citep{Burger2021}. An array of methods have been developed to quantify this similarity, with much of the literature focused on Bayesian dynamic borrowing methods \citep{Yang2023,Damone2024}, propensity score methods \citep{Lim2018,Wang2020}, or a combination of the two \citep{Lin2019,Fu2023}.

For our purposes of trial design, drawing clinical conclusions or real inference is \textit{not} the aim; rather, the simulated trials in this work are used for a power analysis to design a future trial. As explained in Section~\ref{sec:frameworktopowersmart}, we propose a synthetic data generation framework to run in silico trials, where generative models are fit to real pilot SMART data; this is our internal data source. However, pilot trials are typically very small, enrolling only a few dozen participants, thus making it difficult to fit more complex data generation models. To aid with the task of fitting more realistically complex generative models, internal data are augmented with external data, and the models are then fit to the larger, augmented data set. Note that while the external data population may not match the internal population exactly, all trial inclusion and exclusion criteria will be applied to ensure adequate overlap. In the external controls literature, this procedure is also called \textit{static borrowing} \citep{Burger2021}, which we argue is sufficient for our purposes. In this way, we are able to capture a larger sample of the desired population of interest. Since the goal is to plan a future full-scale SMART, it is advantageous to incorporate data that represent a larger sample from the target population, so long as the trial's inclusion and exclusion criteria are met. 

Since EDTRs are specific to SMART designs and hence finding external data that match each treatment randomization stage is likely a difficult task, and because our proposed framework involves a more complex generative model at baseline followed by more simplistic models post-baseline (further explained in Section~\ref{sec:frameworktopowersmart}), we choose to augment only baseline internal data with external information in order to then generate a synthetic trial cohort that ideally represents the population of interest. Practically, this means the external data source should include the same variables as those collected at baseline in the SMART. Ideally, these variables would also be measured or coded in the same manner with the same precision, however we acknowledge that this may not always be the case. Variables that are not shared by both data sources are not included in the augmented data set; variables that exist only in the external data are omitted entirely. In Section~\ref{subsec:simulationmethod} we demonstrate how synthetic data can still be generated for variables that are only in the internal data. Additionally, if the external data do not have the same precision, for example if the external data include categorical variables that are coarser than those in the internal data, then the data augmentation step requires a re-categorization of the internal data such that the given (categorical) variable is coded in a consistent manner in the augmented data. If the external data are finer, then the precision of the internal data is maintained in the augmented data.

\section{Framework to Power SMARTs}
\label{sec:frameworktopowersmart}

Next, we describe the proposed simulation procedure for powering SMARTs for aim 3. We start with a brief description of our previously-proposed synthetic data generation framework that is the backbone of this procedure. We also describe the internal and external data sources used to demonstrate how this methodology works in practice. The simulation framework is then presented in detail, followed by a presentation of the baseline comparator.

\subsection{Synthetic Data Generation Framework}
\label{subsec:syntheticdatagenerationframework}

We previously explored several ways of generating complex, realistic RCT data with one stage of treatment randomization such that the synthetic data distribution closely mimics the real data distribution \citep{Petrakos2025a}. We found that of the methods we compared, a sequential data generation procedure, where the data generation task is split into sequential steps following the temporal ordering of the trial (baseline, treatment, follow-up, final outcome), was far superior to a simultaneous one, where one generative model is fit to all variables. Additionally, as there are many more variables to generate at the first sequential step (baseline) compared to the other steps, the task of generating baseline information is more involved. For this, our investigations showed that an R-vine copula model was most successful at generating (completely-observed) baseline data. Additionally, we found that a simple random draw from a pre-defined probability distribution was the best way to generate synthetic treatment assignment as this mimics a real trial, and that fitting either linear or logistic regression models was sufficient to generate the sequential post treatment randomization variables, including the outcome, where each variable is generated by separate regression models. Generating synthetic data for a continuous variable via linear regression involves three steps: 1) fitting the regression to real data where the model outcome is the variable to be generated and the model predictors are all variables that came before in the trial, 2) predicting synthetic values of this variable on the set of generated synthetic data from previous steps, and 3) inducing an extra source of randomness by sampling from the subset of admissible values defined as the sums of predicted values and model residuals that lie within the support of the variable. This gives the final generated value for each synthetic participant. When using logistic regression to generate a binary variable, the final sampling step is not required, as the predicted synthetic value already reflects variability due to being a random draw from a Bernoulli distribution. This procedure is repeated for each of the subsequent post-randomization variables, ending with the final outcome where the predictors in the generative model are all other variables in the trial. The final synthetic data set is thus the merged set of variables generated at each sequential step: baseline variables, randomized treatment assignment, post-randomization variables, and the outcome. Though the use-cases we examined only involved continuous or binary post-randomization variables, variables of other types can also be generated using this framework as the analyst has the freedom to define the sequential generative models to fit the variable support. This sequential data generation procedure is used in the simulation framework presented in this work.

We further investigated methods to account for the real data missingness mechanism, as missing data are common in RCTs and are rarely missing completely at random \citep{Little2012}, as well as to generate synthetic missingness so as to generate data that are faithful to the underlying generating distribution (in which there is no missingness) as well as to the completely-observed real data distribution \citep{Petrakos2025b}. Our empirical results showed that a multiple imputation (MI) approach was most successful at recovering the real data treatment effect, as compared to complete case and inverse probability weighting (IPW) approaches. As detecting treatment effects is the goal of a power analysis, we implement this approach to handling missing data in this work. Additionally, the available pilot SMART data have a limited sample size, and it has been shown that IPW can be unstable in small samples whereas MI is more stable and thus preferable, especially when modelling longitudinal, partially-observed RCT data where post-randomization variables can be incorporated in the imputation models \citep{Rombach2018}. In our proposed framework to incorporate missingness in the generation of realistic RCT data, MI is incorporated in the sequential data generation framework (R-vine copula at baseline, followed by randomized treatment allocation, then sequential regression models) by multiply-imputing the real trial data, fitting the sequential regression models to each of the $m$ imputed data sets, generating $m$ versions of the synthetic variable, and finally sampling from the $m$ versions so one final version of the synthetic data is retained. Since here we assume no missingness at baseline, the imputed data are only relevant in generating post-baseline variables. This previously-proposed framework also allows for the generation of synthetic missingness by modelling the probability of being observed at each sequential generation step using logistic regression (as is standard in an inverse probability weighting approach to handling missing data). Missingness is then imposed on the synthetic data based on the synthetic probabilities of being observed. It is possible to return either a completely-observed synthetic data set or one that is partially-observed. As it is more realistic to estimate the power of a trial design while also accounting for participant drop-out in a more sophisticated manner than simply a fixed inflation factor, the framework we propose here for powering SMARTs utilizes partially-observed synthetic data, where the proportion of drop-out at each stage is informed by the pilot trial data.

\subsection{Data}
\label{subsec:data}

\begin{figure}[ht]
    \centering
    \includegraphics[width=\linewidth]{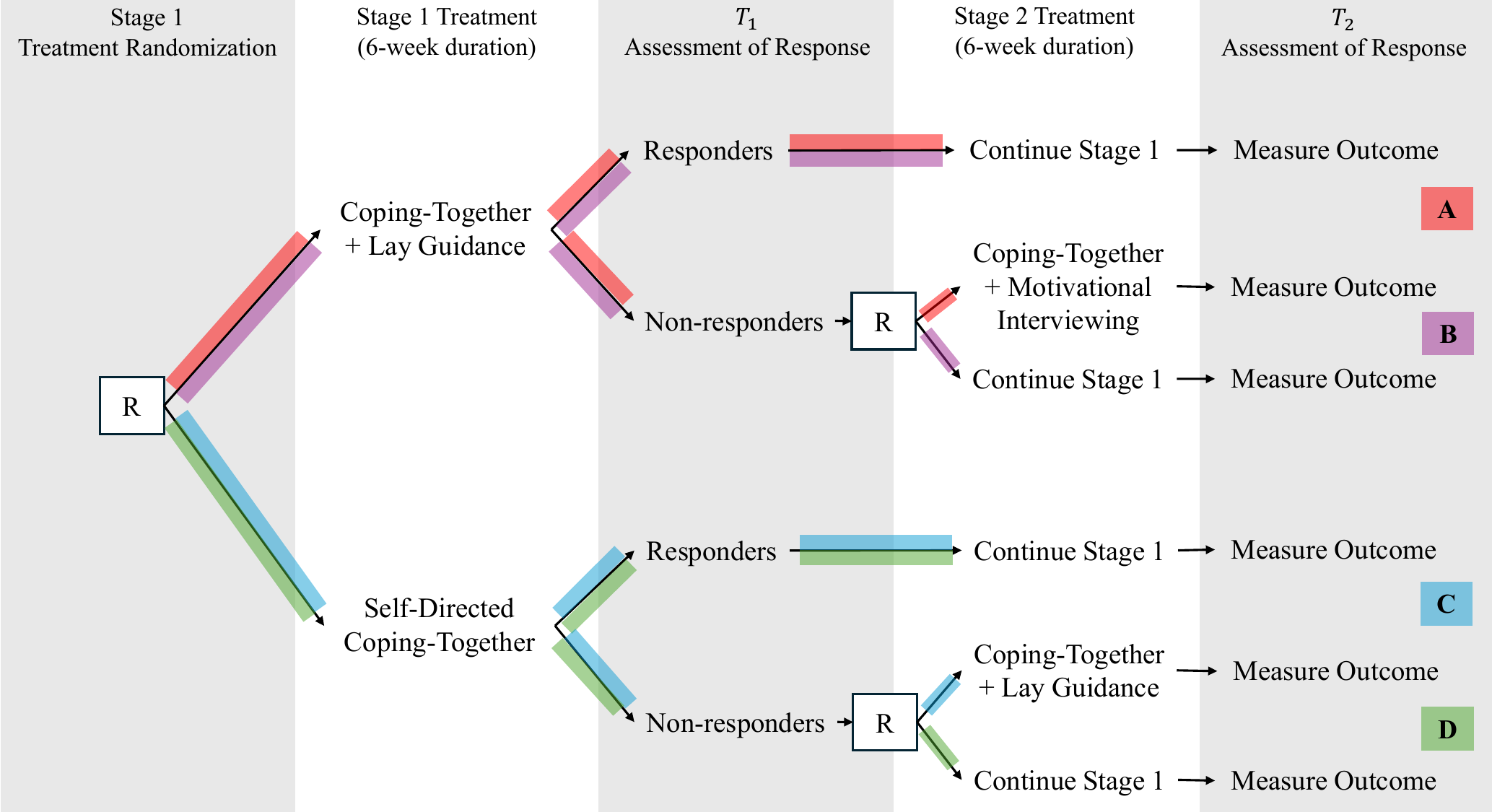}
    \caption{Flowchart showing the pilot cancer SMART design, with labels for each of the four potential strategies (Strategy A in red, B in purple, C in blue, D in green). The symbols R in boxes represent points of treatment (re-)randomization.}
    \label{fig:smartflowchart}
\end{figure}

To demonstrate how the proposed framework operates, we use two example data sets -- one for the internal SMART and the other for the external source. Internal data are from a pilot SMART that investigated self-management interventions to handle physical and psychosocial concerns among cancer patients and their caregivers \citep{Lambert2025}. Though pairs of patients and caregivers were enrolled in the study, we focus on the patients and omit caregiver data. Available data are from 48 patient participants, recruited across three Canadian cancer centers from 2021 to 2022. Inclusion criteria are Stage 0-III cancer diagnosis, cancer treatment in the prior twelve months, and at least low anxiety measured as a Distress Thermometer (DT) score of at least four. Baseline data include age, sex (female, male), language (English, French), marital status (married, common law, separated, divorced, widowed), education level (high school, post-secondary, university), country of birth (Canada, other), DT score, and anxiety score measured by the 7-item hospital and anxiety depression scale (HADS) \citep{Zigmond1983}. The exposure of interest is the assigned treatment sequence. At stage 1, participants were randomized by a stratified (by baseline HADS), block-randomized design with random block sizes of two or four. Only non-responders were re-randomized, whereas responders remained on stage 1 treatment, thus resulting in four strategies labelled A -- D: A) begin with Coping-Together + Lay Guidance, if responder remain with Coping-Together + Lay Guidance, if non-responder assign Coping-Together + Motivational Interviewing; B) begin with Coping-Together + Lay Guidance, if responder remain with Coping-Together + Lay Guidance, if non-responder remain with Coping-Together + Lay Guidance; C) begin with Self-Directed Coping-Together, if responder remain with Self-Directed Coping-Together, if non-responder assign Coping-Together + Lay Guidance; and D) begin with Self-Directed Coping-Together, if responder remain with Self-Directed Coping-Together, if non-responder remain with Self-Directed Coping-Together. This design is depicted in Figure~\ref{fig:smartflowchart}. At $T_1$ (six weeks after baseline), HADS score, DT score, and responder status were recorded. As noted above, the original pilot study enrolled patient/caregiver dyads and responder status was determined as a function of both participant scores. For the purpose of this work, focusing only on the patient participant, a participant was considered a responder if either baseline DT $\geq5$ and $T_1$ DT $<$ baseline DT, or if baseline DT $<5$ and $T_1$ DT $<2$ $+$ baseline DT (i.e., those with baseline DT $<5$ were still a responder if DT increased by at most 1). See the Supplementary Materials for the original pilot study, dyadic definition of responder status. For simplicity, we focus on only one of the outcomes, HADS, which was measured at $T_2$ (twelve weeks after baseline). Missing sociodemographic data were minimal at baseline, as education is the only variable with missingness (2\%, 1 participant). 6\% and 7\% of participants dropped out at $T_1$ and $T_2$, respectively.

The external data source is an RCT that compared a novel telephone-delivered depression self-care intervention, called CanDirect, to usual care among cancer patients \citep{Lambert2021}. Inclusion criteria are at least 18 years of age, completed cancer treatment in the previous ten years and in remission, and mild to moderate depressive symptoms as measured by a score of 8-19 on the Patient Health Questionnaire-9 \citep{Kroenke2001,Thekkumpurath2011}. Participants with metastatic disease, suicidal intent, moderate to severe cognitive issues, who did not understand English or French, or who were already receiving psychological treatment or had begun a new or adjusted antidepressant medication in the six weeks prior to the start of the trial were excluded. The exposure of interest is supported self-care interventions (CanDirect) plus usual care; the comparator is usual care only. At baseline, the following variables were recorded for the 245 study participants: age, sex (female, male), language (English, French), marital status (single or never married, married or common law, separate or divorced, widowed), education level (less than high school, completed high school, post-secondary, university), country of birth (Canada, other), and HADS score. Missingness at baseline is low, with marital status and country of birth having the highest proportions of missingness (1\%); 97\% of participants have completely observed baseline data. Since the external data source is only pertinent to augment internal baseline data within the context of this work, variables collected after baseline are not included. Additionally, marital status and education level are re-categorized during the data augmentation step as these variables are coded differently in the internal and external data. Participants in the RCT with missing values are not included in the augmented baseline data due to the very low proportion of missingness. Though our external data source example is an RCT, observational data can also be utilized, so long as the trial inclusion and exclusion criteria are met (see Section~\ref{sec:dataaugmentation}).

Thus, the baseline data $\mathbf{X}$ include age (continuous, in years), sex (male, female), marital status (single, not single), language (English, French), country of birth (Canada, other), education level (high school or less, post-secondary, university), HADS (continuous, 0-21), and DT (continuous, 0-10). We denote the baseline data shared by both data sources, which form the augmented baseline data set, as $\mathbf{X}_{-1}$ and the baseline data available only in the internal data (a single variable in this case, DT) as $X_1$. Hence $\mathbf{X}_{-1}$ and $X_1$ together form $\mathbf{X}$. $A_1=\{0,1\}$ with $0$ representing Self-Directed Coping-Together and $1$ representing Coping-Together + Lay Guidance. The variables collected at $T_1$ are $Z_{11}$ (DT score at $T_1$), $Z_{12}$ (HADS score at $T_1$), and $R_1=\{0,1\}$ where $0$ is for non-responder and $1$ is for responder. $A_2=\{0,1,2\}$ where $A_2=2$ represents Coping-Together + Motivational Interviewing, and the levels 0 and 1 are as defined in $A_1$. The outcome $Y$ is HADS at $T_2$. 

\subsection{Simulation Methodology}
\label{subsec:simulationmethod}

The premise of the proposed simulation approach to powering a SMART is to generate \textit{realistic} trial data that one would expect to observe under a given design. Hence, we extend the framework described in Section~\ref{subsec:syntheticdatagenerationframework} to generate SMART data, with the key modifications being to generate synthetic data for variables in the internal pilot SMART baseline data that are not available in the external data, incorporate more than one stage of treatment randomization, generate multiple variables (e.g., tailoring variables) at each sequential step, and generate responder status (as this is a key feature of SMARTs). These steps are described in further detail in Section~\ref{subsubsec:generatesyntheticSMARTdata}. As our example SMART involves only two stages, the following descriptions involve two stages of treatment randomization, though the number of stages could be greater than two and the framework could be extended to incorporate these additional steps without any difficulty.

Once equipped with a synthetic data set representing a potential SMART design, one can perform a hypothesis test. Since our purpose is to power SMARTs to address aim 3 (comparing EDTRs that differ in stage 1 treatment assignment), a two-sample t-test suffices as there is no overlap in the subgroups (with significance level set to 0.05 in all subsequent simulations). As in other simulation-based power analyses, this process is then repeated $N$ times, and power is estimated to be the proportion of correct null hypothesis rejections out of $N$ \citep{Jeffers2024}. Here, the critical word is \textit{correct}, meaning it is necessary for the simulated data to represent a true effect size for which researchers aim to have the power to detect in their trial design. The manner in which this true effect size is ensured in our simulations is explained in Section~\ref{subsubsec:effectsize}.

As it is of interest to determine the minimally-sufficient sample size to detect the effect size of interest, this whole procedure is then repeated across a range of sample sizes, $n$. Since simulation-based methods are more time and resource intensive than sample size calculators, a coarser grid of $n$ is first defined to give an approximate required sample size, though the entire procedure could then be repeated across a finer grid within a narrower range if a more precise value of $n$ is desired. In the simulations presented here, $n=\{100, 200, 300,...,3000\}$. For illustrative purposes, we demonstrate how to employ this framework to power a SMART for one EDTR comparison, which we arbitrarily choose to be Strategy A versus Strategy C (refer to Figure~\ref{fig:smartflowchart}). However, it is common for a SMART to have multiple study aims, in which case this framework could be implemented for each aim 3 comparison, where the final recommended minimal sample size for the future full-scale SMART would be the maximum of the estimated sample sizes.

The utility of this framework is to power a full-scale trial after a pilot trial has been completed. A natural order of events when designing the full-scale trial is: determine the necessary sample size for a full-scale SMART with the same design as the pilot, then investigate whether an alternate design would lead to an increase in observed effect size and thus a decrease in required sample size. If so, then it would be of interest to calculate the sample size required for the competing design that would likely result in the largest effect size. Here, we make the distinction between a ``fixed" effect size and an ``observed" effect size. The former is the approach when using a sample size calculator or simple formulae where the analyst assumes or makes an educated guess about what effect size is likely to be observed (or what they hope to observe or be able to detect) in the future trial. In contrast, our proposed method acknowledges that the observed effect size when running the trial is very unlikely to be exactly the same as what is set in the calculator or formulae, and thus this uncertainty is incorporated in our simulation-based sample size estimation. Thus, we first illustrate the proposed framework through the implementation of the original pilot SMART design, which we also set to be the first competing design. The other competing designs and further investigations regarding gains in effect size and subsequent sample size calculations for competing designs are defined in Section~\ref{subsec:competingdesigns}.

\subsubsection{Generate Synthetic SMART Data with Missingness}
\label{subsubsec:generatesyntheticSMARTdata}

To generate synthetic data, which we denote by a superscript $s$, generative models are sequentially fit to the real data, denoted by a superscript $r$. The first step is to generate $\textbf{X}^s$. Using the framework from Section~\ref{subsec:syntheticdatagenerationframework}, this involves fitting an R-vine copula to augmented $\textbf{X}^r_{-1}$, as defined in Section~\ref{subsec:data}. The fitted copula is then used to generate synthetic versions of the variables in the augmented data set, denoted $\textbf{X}^s_{-1}$; these are the variables that are shared by the internal and external data sources. To generate $X_1^s$, a separate regression model is fit to the internal pilot SMART data only, with $X_1^r$ as the outcome and all other baseline variables as the predictors. Thus, this variable is generated separately from $\textbf{X}^s_{-1}$, following the steps outlined previously when using regression to generate variables one at a time. As this model is fit to only the internal SMART data, the sample size is much smaller than that of the augmented data set, which is a potential limitation for all subsequent generative models fit only to the pilot SMART data. The implication of this is further discussed in Section ~\ref{sec:discussion}. The vector $X^s_1$ is then merged with $\textbf{X}^s_{-1}$, resulting in the synthetic baseline cohort, $\textbf{X}^s$.

The next step is to generate $A_1^s$, which in our motivating context follows the same stratified block randomized design as the original pilot SMART, with participants stratified by baseline HADS: $<8$, $[8-10]$, $[11,21]$. Here, no model is fit to the real data. (Note that in a more general setting, randomization may be simple or unstratified.)

Following stage 1 treatment is $T_1$, where participant drop-out begins to occur, leading to missing values at $T_1$, $A_2$, and $T_2$ and thus further limiting the available data for fitting generative models post-randomization. As mentioned in Section~\ref{subsec:syntheticdatagenerationframework}, our solution is to utilize MI. The internal SMART data are imputed $m$ times, where all variables in the trial are included in the imputation models. Note, however, that $R^r_1$ has a deterministic relationship with $X^r_1$ and $Z^r_{11}$, and since $A^r_2$ is partially determined by $A^r_1$ and $R^r_1$, passive imputation is employed for $R^r_1$ and $A^r_2$ when implementing MI through mice to ensure these deterministic relationships are maintained in the imputed data sets \citep{vanBuuren2011}. Following this, the generative models described below for variables post-stage 1 randomization are fit to each imputed data set, thus leading to $m$ synthetic sets of each variable. To attain one final synthetic vector per variable, a sampling procedure is utilized where the $i^{\text{th}}$ observation of a given variable is randomly sampled from the set of all $i^{\text{th}}$ observations of this variable from the $m$ imputed data sets. More details regarding this procedure can be found in \cite{Petrakos2025b}.

We continue with defining the generative models for variables following stage 1 treatment randomization. Next are the variables measured at $T_1$: $Z_{11}^s$, $Z_{12}^s$, $R_1^s$. Again, separate regression models are fit to generate $Z_{11}^s$ and $Z_{12}^s$, where the outcome for each model is either $Z_{11}^r$ or $Z_{12}^r$ (depending on which variable is being generated) and the predictors are all variables that are temporally precedent: $\textbf{X}^r$, $A_1^r$. In our example, $Z_{11}^r$ and $Z_{12}^r$ are continuous variables and thus are generated using linear regression; $Z_{11}^s$ and $Z_{12}^s$ are then generated by first predicting values for the set $\{\textbf{X}^s,A_1^s\}$, then sampling from the set of admissible values. For $Z_{11}^s$, this set is $\{0 \leq \text{prediction} + \text{residuals} \leq 10\}$, and for $Z_{12}^s$, this set is $\{0 \leq \text{prediction} + \text{residuals} \leq 21\}$, as DT and HADS can only take values within these two respective ranges. Note that the set of predictors included in generative models for variables collected at $T_1$ do not include other variables collected at the same time point. For instance, $Z^r_{11}$ is not a predictor in the model to generate $Z^s_{12}$. Because of this, the choice of which variable to generate first is arbitrary, however the order would matter if the analyst instead defines a generative model that incorporates variables collected at the same stage in the trial. Implications of this choice are further discussed in Section~\ref{sec:discussion}. Next, responder status in the pilot is defined by a deterministic function involving $X^r_1$ and $Z^r_{11}$ and thus is applied in the synthetic data to generate $R_1^s$: if $X^s_1\geq5$, and $Z^s_{11}<X^s_1$, then $R^s_1=1$; if $X^s_1<5$, and $Z^s_{11}<X^s_1+2$, then $R^s_1=1$; else $R^s_1=0$. Similar to generating $A_1^s$, no model is fit to the real data to generate $R^s_1$.

The next time point in the SMART is stage 2 treatment assignment. To generate $A^s_2$, a simple random treatment assignment is performed by drawing from a Bernoulli$(0.5)$ as there are only two potential treatment arms. However, the trial design necessitates that only the non-responders are re-randomized, and the two arms at stage 2 differ depending on assigned treatment at stage 1. Thus, $A^s_2$ is generated via the following: if $R^s_1=1$, then $A^s_2=a^s_1$ where $a^s_1$ is the assigned $A^s_1$ treatment; if $R^s_1=0$ and $A^s_1=1$, then assign one of $A^s_2\in\{1,2\}$ based on a random Bernoulli draw; if $R^s_1=0$ and $A^s_1=0$, then assign one of $A^s_2\in\{0,1\}$ based on a random Bernoulli draw.

Finally, $Y$ is measured in the trial. Normally, $Y^s$ would be generated by fitting another regression model (here, $Y$ is continuous HADS and thus the generative model is linear regression), predicting on the set of synthetic data, and then inducing additional randomness. This would result in synthetic data with a distribution that closely mimics that of the real data. However, recall that to estimate power, the synthetic data must represent a known (standardized) effect size, $\delta$. We thus modify slightly the manner in which $Y^s$ is generated to ensure $\delta$ is at the desired level, as described in Section~\ref{subsubsec:effectsize}. 

The final step is to impose missingness in the synthetic data to represent attrition. As we assume completely-observed data at baseline and stage 1 treatment, synthetic participants are randomly selected to drop-out at $T_1$ and $T_2$ with probability equal to the observed proportion of drop-out at each stage in the pilot trial. Ideally, a separate generative model would be fit instead, modelling either a) the probability of being observed at the given time point and then used to predict the synthetic probability of being observed, or b) a multi-level categorical variable representing lost to follow-up, with levels being not lost to follow-up, dropped out at $T_1$, dropped out at $T_2$ with predictors being all other variables collected during the trial. Due to sample size constraints with the available pilot data, very few participants dropped out (three at $T_1$, three at $T_2$) and hence it was not possible to fit either model (though, sample size allowing, this would be our recommended method for implementing participant attrition when running these trials in simulation). This results in partially-observed synthetic data including those who drop-out at $T_1$ ($\{\tilde{Z}^s_{11},\tilde{Z}^s_{12},\tilde{R}^s_{1},\tilde{A}^s_2\}$) and at $T_2$ ($\tilde{Y}^s$).

Thus, the final returned synthetic sample with participant attrition is $\{\textbf{X}^s,A^s_1,\tilde{Z}^s_{11},\tilde{Z}^s_{12},\tilde{R}^s_{1},\tilde{A}^s_2,\tilde{Y}^s\}$. Of course, fully-observed synthetic SMART data could be returned instead if desired, such as if no attrition is to be expected: $\{\textbf{X}^s,A^s_1,Z^s_{11},Z^s_{12},R^s_{1},A^s_2,Y^s\}$.

\subsubsection{Effect Size}
\label{subsubsec:effectsize}

For pedagogical purposes, we focus on the comparison of Strategy A and Strategy C, shown in red and blue, respectively, in Figure~\ref{fig:smartflowchart}. The derivations for the remaining EDTR comparisons can be found in the supplementary materials. Specifically, Strategy A imposes $A_1=1$, and $A_2=1$ for responders and $A_2=2$ for non-responders. Strategy C imposes $A_1=0$, and $A_2=0$ for responders and $A_2=1$ for non-responders. Note that $A_2=1$ is a possible treatment assignment in both strategies. 

To estimate power, we need to ensure that there is a true, known, treatment effect to detect in the synthetic data so that it is clear for what $\delta$ the design is powered to detect. We fix a known $\delta$ in modifying the generation of $Y^s$, though the final sample size calculation takes into account that the generated synthetic data will have an observed $\delta$ that is different, though close, to the fixed quantity of $\delta$. The choice of this fixed quantity of $\delta$ can be motivated by the values that were assumed in the pilot study.

First, recall the definition of $\delta$ for aim 3 \citep{Oetting2007}: 
\begin{equation}
\label{eq:deltadefrearrangeAvC}
    \delta=\frac{1}{S}\left( \mathbb{E}\left[ Y|A_1=1,A_2=a^{\text{Strat A}}_2 \right] - \mathbb{E}\left[ Y|A_1=0,A_2=a^{\text{Strat C}}_2 \right] \right)
\end{equation}
where, for notional convenience, $S:=\sqrt{\frac{\text{Var}\left[ Y|A_1=1,A_2=a^{\text{Strat A}}_2\right]+\text{Var}\left[ Y|A_1=0,A_2=a^{\text{Strat C}}_2\right]}{2}}$. 
$A_2=a^{\text{Strat A}}_2$ denotes the assigned treatment at stage 2 under Strategy A ($a_2=1$ for responders and $a_2=2$ for non-responders), and $A_2=a^{\text{Strat C}}_2$ denotes the assigned treatment at stage 2 under Strategy C ($a_2=0$ for responders and $a_2=1$ for non-responders). 

In the synthetic data generation framework, the outcome $Y$ is modelled as 
\begin{equation}
\label{eq:outcomemodel}
    Y=\beta_0+\beta_1\mathbf{X} + \beta_2A_1+\beta_3\mathds{1}(A_2=1)+\beta_4\mathds{1}(A_2=2)+\epsilon \text{ , with } \epsilon \sim N(0,\sigma^2)
\end{equation}
where, without loss of generality, $\mathbf{X}$ are participant covariates (baseline variables and stage 1 tailoring variables). We drop the $r$ and $s$ superscripts for notational convenience, since the generative model for the outcome applies to both the real and synthetic data (it is fit to the real data and used to predict the synthetic data). Then the conditional expectation of the outcome for each strategy can be estimated using the outcome model and further conditioning on the response status:
\begin{equation}
\label{eq:outcomeconditionalexpectationdecomp}
    \begin{split}
        \mathbb{E}\left[Y|A_1,A_2\right]&=\mathbb{E}\left[\mathbb{E}\left[Y|A_1,A_2,R_1\right]\right] \\
        &=\sum_{r_1\in\{0,1\}}\mathbb{E}\left[Y|A_1,A_2,R_1=r_1\right]\cdot \text{Pr}\left(R_1=r_1|A_1,A_2\right) \\
        &= \sum_{r_1\in\{0,1\}}\mathbb{E}\left[Y|A_1,A_2,R_1=r_1\right]\cdot \text{Pr}\left(R_1=r_1|A_1\right)
    \end{split}
\end{equation}
where $r_1=0$ denotes a non-responder and $r_1=1$ denotes a responder to stage 1 treatment. $R_1$ only depends on $A_1$ as $A_2$ occurs after responder status is defined. Notably, we do not impose the assumption that response rate is the same across stage 1 treatment groups.

Then based on Equation~\ref{eq:outcomeconditionalexpectationdecomp} and denoting $p_1$ the response rate (proportion of responders) to $A_1=1$ and $p_0$ the response rate to $A_1=0$,
\begin{equation*}
    \begin{split}
        \mathbb{E}\left[Y|A_1=1,A_2=a^{\text{Strat A}}_2\right]&=\mathbb{E}\left[Y|A_1=1,A_2=a^{\text{Strat A}}_2,R_1=0\right]\cdot \text{Pr}\left(R_1=0|A_1=1\right) \\ 
        &\hspace{1cm}+\mathbb{E}\left[Y|A_1=1,A_2=a^{\text{Strat A}}_2,R_1=1\right]\cdot \text{Pr}\left(R_1=1|A_1=1\right) \\
        &=\left(\beta_0+\beta_1\mathbb{E}[\mathbf{X}]+\beta_2\cdot1+\beta_3\cdot0+\beta_4\cdot1\right)\cdot\left(1-p_1\right) \\
        &\hspace{1cm}+ \left(\beta_0+\beta_1\mathbb{E}[\mathbf{X}]+\beta_2\cdot1+\beta_3\cdot1+\beta_4\cdot0\right)\cdot(p_1)\\
        &=\beta_0+\beta_1\mathbb{E}[\mathbf{X}]+\beta_2+\beta_3\left(p_1\right)+\beta_4\left(1-p_1\right)
    \end{split}
\end{equation*}
and
\begin{equation*}
    \begin{split}
        \mathbb{E}\left[Y|A_1=0,A_2=a^{\text{Strat C}}_2\right]&=\mathbb{E}\left[Y|A_1=0,A_2=a^{\text{Strat C}}_2,R_1=0\right]\cdot \text{Pr}\left(R_1=0|A_1=0\right) \\ 
        &\hspace{1cm}+\mathbb{E}\left[Y|A_1=0,A_2=a^{\text{Strat C}}_2,R_1=1\right]\cdot \text{Pr}\left(R_1=1|A_1=0\right) \\
        &=\left(\beta_0+\beta_1\mathbb{E}[\mathbf{X}]+\beta_2\cdot0+\beta_3\cdot1+\beta_4\cdot0\right)\cdot\left(1-p_0\right) \\
        &\hspace{1cm}+\left(\beta_0+\beta_1\mathbb{E}[\mathbf{X}]+\beta_2\cdot0+\beta_3\cdot0+\beta_4\cdot0\right)\cdot(p_0) \\
        &=\beta_0+\beta_1\mathbb{E}[\mathbf{X}]+\beta_3\left(1-p_0\right).
    \end{split}
\end{equation*}
Subtracting these two conditional expectations and plugging into Equation~\ref{eq:deltadefrearrangeAvC} gives
\begin{equation}
\label{eq:deltabetasrelationship}
    \delta\cdot S=\mathbb{E}\left[Y|A_1=1,A_2=a^{\text{Strat A}}_2 \right]-\mathbb{E}\left[ Y|A_1=0,A_2=a^{\text{Strat C}}_2\right]=\beta_2+\beta_3\left(p_1+p_0-1\right)+\beta_4\left(1-p_1\right).
\end{equation}

Then, to ensure a data-generating standardized effect size of $\delta$ in the synthetic data (such that the power calculation is valid), select $\beta_2,\beta_3,\beta_4$ such that the above equality holds. This requires fixing the values of $S$, $p_0$, $p_1$. We choose to set $S=\hat{S}^r$ as the empirical estimate from the real pilot data, whereas $p_0=\hat{p}^s_0,p_1=\hat{p}^s_1$ are set as the empirical estimates from the synthetic data following the responder status definition of the original pilot design. Other choices such as expert opinion could also be used. $\delta$ is pre-defined as the desired standardized effect size (e.g., 0.2). 

There are multiple ways to define the $\beta$ parameters. We describe one such way in this paper, though other solutions are also possible. Let $\beta_2=\hat{\beta}_2$ and $\beta_3=\hat{\beta}_3$, where $\hat{\beta}_2,\hat{\beta}_3$ are estimates from the fitted outcome generation model (fitted to pilot data). Then solve for $\hat{\beta}_4$: $\hat{\beta}_4=\frac{1}{1-\hat{p}^s_1}\left[\delta \cdot \hat{S}^r-\hat{\beta}_2-\hat{\beta}_3\left(\hat{p}^s_1+\hat{p}^s_0-1 \right) \right]$.

With $\beta_2,\beta_3,\beta_4$ fixed so as to achieve the desired effect size, the remaining $\beta$ parameters for the outcome model can be set to match those from a fitted model based on the pilot SMART data and then used to predict and finally generate the synthetic outcome using the synthetic data $\{\textbf{X}^s,\allowbreak A^s_1,\allowbreak Z^s_{11},\allowbreak Z^s_{12},\allowbreak R^s_{1},\allowbreak A^s_2,\allowbreak Y^s\}$ (where prediction and final generation are two separate steps, as outlined in Section~\ref{subsec:syntheticdatagenerationframework}). This ensures that in each simulation run and for each synthetically-generated data set representing a given study design, there is a ``true" data generating effect size between the two strategies of interest. Note that the quantities $\hat{p}^s_0,\hat{p}^s_1$ change per simulation, as a different synthetic data set is generated each run. Then, the power calculation follows as the proportion of null hypothesis rejections out of the $N$ simulation runs. 

\subsection{Competing Designs}
\label{subsec:competingdesigns}

The procedure outlined above fixes a single data generating mechanism using the original pilot data under the original pilot study design. With this, a power calculation for varying $n$'s can be performed for the original study design, which is useful for powering a future full-scale SMART with the same design operating characteristics as the pilot. Of course, it may also be of interest to investigate alternative decisions in the trial design; for example, we shall consider whether re-defining responder status may lead to a larger observed effect size, and thus gains in power and minimizing sample size requirements.

First, competing designs should be defined prior to running the simulations, and prior to running the full scale trial. In this paper, we use responder status as the study design characteristic that we vary to define competing designs. Recall that in the original pilot, henceforth referred to as Design I, responder status is defined in the following way: if baseline DT is at least 5, and stage 1 DT is less than baseline DT, then the participant is a responder; if baseline DT is less than 5, and DT decreased, stayed the same, or increased by no more than 1 point between baseline and stage 1, then the participant is a responder; else they are a non-responder. This design resulted in $\hat{p}^{r_{I}}_1=0.52$ and $\hat{p}^{r_{I}}_0=0.71$ in the original pilot data, where the superscript $r_{I}$ indicates the real data with responders defined according to Design I. We thus define an alternative design with a stricter, still clinically-driven responder status definition, giving smaller probabilities of responding to $A_1$ and thus larger proportions of participants re-randomized at stage 2 (Design II), as well as a design with a purely data-driven definition of responder status that better ensures balance between responders and non-responders (Design III). Specifically, Design II differs from Design I in that it defines responder status as: if both DT and HADS decrease by at least 2 points between baseline and stage 1, then the participant is a responder; else they are a non-responder. Re-assigning responder status to the real data participants using Design II's definition gives $\hat{p}^{r_{II}}_1=0.13$ and $\hat{p}^{r_{II}}_0=0.33$ (estimates from the real data are calculated by redefining real pilot participants according to the responder status definition in Design II). Design III implements the following data-driven definition of responder status: if stage 1 DT is less than the treatment subgroup's median DT in the pilot, then the participant is a responder; else they are a non-responder. Under this design, reassignment of responder status using the real pilot data gives $\hat{p}^{r_{III}}_1=0.65$ and $\hat{p}^{r_{III}}_0=0.5$. Note that $\hat{p}^{r_{III}}_1$ is not exactly $0.5$ due to participant drop-out at stage 1.

Now, we are interested in fixing the data generating mechanism under the original pilot design (Design I) and investigating whether the pre-specified competing designs lead to potential gains in effect size. The implementation procedure applies to each competing design. Synthetic data are generated as described before, though now three versions of $R^s_1$ are generated, one per design: $R^{s_{I}}_1,R^{s_{II}}_1,R^{s_{III}}_1$. This results in $\{\hat{p}^{s_{I}}_0,\hat{p}^{s_{I}}_1\},\{\hat{p}^{s_{II}}_0,\hat{p}^{s_{II}}_1\},\{\hat{p}^{s_{III}}_0,\hat{p}^{s_{III}}_1\}$. Then, three versions of $A^s_2$ are simulated, again one per design: $A^{s_{I}}_2,A^{s_{II}}_2,A^{s_{III}}_2$. Then setting $\delta=0.2$ (a choice informed by the benchmark value chosen in the original pilot SMART), fix $\beta$ parameters in the outcome generating model (Equation~\ref{eq:outcomemodel}) as the fitted values (per $m$ imputed data set) and set $\hat{\beta}_4$ as the value derived in Equation~\ref{eq:deltabetasrelationship} using $\hat{p}^{s_{I}}_0,\hat{p}^{s_{I}}_1$ from Design I. This gives $m$ sets of $\hat{\beta}$ (recall that due to missingness in the real pilot data, this procedure of generating $Y^s$ is performed $m$ times). Hence, there is a distribution of $\hat{\beta}$ values that represents the ``true", and fixed, data generating mechanism. Now fixing these $\beta_0,\beta_1,\beta_2,\beta_3$ as the fitted values and $\beta_4$ as $\hat{\beta}_4$ calculated from Equation~\ref{eq:deltabetasrelationship}, calculate the value of $\delta$ for Design II and Design III using Equation~\ref{eq:deltabetasrelationship} with $p_1,p_0$ set as $\hat{p}^{s_{II}}_0,\hat{p}^{s_{II}}_1$ and $,\hat{p}^{s_{III}}_0,\hat{p}^{s_{III}}_1$, respectively. This results in a distribution of $\delta$ estimates per design across the $m$ imputed data sets. These distributions of $\delta$ represent the effect size that one would expect to observe under each competing design, assuming a fixed data generating mechanism (derived from Design I). Here, $m$ is set to 1000 to adequately account for uncertainty in the imputation procedure and $n$ is set to 10,000 to ensure a sufficiently-large data set. 

Then, the next step is to identify which competing design, if any, leads to an increase in effect size and to calculate the required sample size under this alternate design. In plotting the distributions of $\delta$, the analyst can visually identify which distribution falls furthest to the right of the chosen value of $\delta$ when fixing the data generating mechanism (0.2 in this example). Once identified, the sample size for this design can be estimated using the \textbf{fixed} data generating mechanism (i.e., the set of $\beta$ parameters fixed under Design I). The simulation procedure continues by generating the version of $Y^s$ that corresponds to the given design, using Equation~\ref{eq:outcomemodel} and plugging in the version of the synthetic data corresponding to the chosen competing design. As before, a hypothesis test is performed and the sample size is estimated as the proportion of correct null hypothesis rejections out of $N=1000$. 

\subsection{Comparator}
As mentioned in Section~\ref{sec:intro}, to our knowledge, no simulation methods currently exist in the literature for powering SMARTs to study aim 3. However, arguably one of the most straightforward and user-friendly existing tools to power a SMART for aim 3 is a sample size calculator, and as our example SMART has a continuous outcome, we employ SMARTsize as our comparator \citep{Seewald2024}. This calculator requires the input of the following features: outcome (binary or continuous), primary aim (1, 2, or 3), EDTR comparison (which two strategies to compare), probability of response to $A_1$, denoted $p$ (where it is assumed $p:=p_1=p_0$), $\delta$, type of hypothesis test (one- or two-sided), $\alpha$, and $1-\beta$. Since the calculator assumes that response rate is the same across stage 1 treatment groups, to perform comparisons, $p$ is set as the mean of $\hat{p}^r_1$ and $\hat{p}^r_0$ for each design. To account for participant attrition, we inflate the sample size estimates from the calculator by a factor equal to the overall observed proportion of drop-out from the pilot trial, which was approximately 13\%. In the case of a decimal value, the final sample size is then rounded up to the nearest integer.

The sample size calculator effectively assumes a different data generating mechanism per design (as $p$ is the only calculator input that changes across designs, in our illustrative example), and it calculates the sample size per design (i.e., per $p$) across varying levels of $\delta$. However, in the procedure outlined above (Section~\ref{subsec:competingdesigns}), a fixed data generating mechanism is assumed. Hence, we propose a second way of calculating the required sample size where the data generating mechanism varies per design and instead $\delta$ is fixed, thus providing a direct comparison with the results from SMARTsize. For the simulation procedure, the only difference for this alternative approach is the calculation of $\hat{\beta}_4$, where now the values chosen for $p_0,p_1$ are the estimates from the synthetic data \textbf{generated under each design}. For instance, when performing the power analysis for Design II, $\hat{\beta}_4=\frac{1}{1-\hat{p}^{s_{II}}_1}\left[\delta \cdot \hat{S}^r-\hat{\beta}_2-\hat{\beta}_3\left(\hat{p}^{s_{II}}_1+\hat{p}^{s_{II}}_0-1 \right) \right]$. The rest follows as described in Section~\ref{subsec:simulationmethod} for each of the competing designs, where $A^{s_{II}}_2$ and $A^{s_{III}}_2$ are used in the prediction of $Y^{s_{II}}$ and $Y^{s_{III}}$, respectively. 

\section{Results}
\label{sec:results}

\begin{figure}[H]
    \centering
    \includegraphics[width=\linewidth]{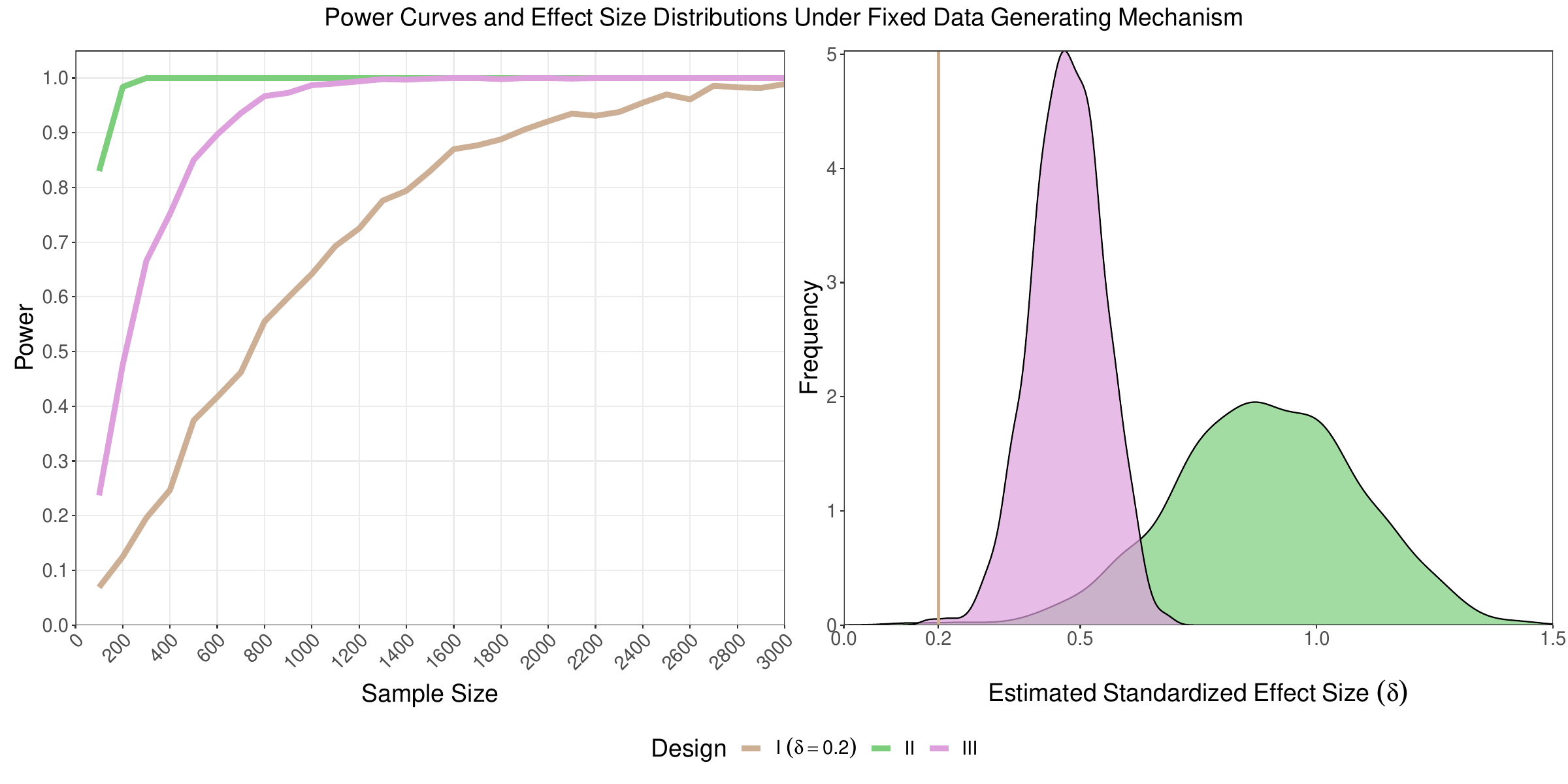}
    \caption{On the left: power curves for Design I (the original pilot design) as well as the two competing designs (Design II in green, Design III in purple), under a fixed data generating mechanism (setting $\delta=0.2$ under Design I). On the right: distribution of computed values of $\delta$ using Equation~\ref{eq:deltabetasrelationship}, setting $\beta$ parameters in the outcome generation model for Design II and Design III to be equal to the values of $\beta$ under Design I for which $\delta\approx0.2$.}
    \label{fig:powercurvedeltadist}
\end{figure}

Presented are the results for both simulation investigations (first fixing the data generating mechanism and varying $\delta$, then fixing $\delta$ and varying the data generating mechanism). Parallelized simulations were run using R version 4.3.1 on a machine with 16 GB RAM and 12 cores. Figure~\ref{fig:powercurvedeltadist} shows the results from the former investigation, where the data generating mechanism is fixed under Design I and the expected effect size and subsequent sample size under competing designs are estimated. Here, we see a substantially higher $\delta$ under the competing designs, assuming the same data generating mechanism, with the largest gain occurring under Design II as evidenced by the $\delta$ distribution for Design II being furthest to the right. This also aligns with the sample size results (left hand side of Figure~\ref{fig:powercurvedeltadist}), where Design II requires a much smaller sample size to achieve the same level of power, assuming the same data generating mechanism.

\sloppy In implementing these simulations, $\hat{p}_1$ seemed to drive the observed effect size in the synthetic data. In this example, where Strategy A is compared to Strategy C, $\delta/S=\mathbb{E}\left[Y|A_1=1,A_2=a^{\text{Strat A}}_2 \right]-\mathbb{E}\left[ Y|A_1=0,A_2=a^{\text{Strat C}}_2\right]$ is large when there is a higher proportion of participants randomized to treatment $A_2=2$ (Coping-Together + Motivational Interviewing), since $A_2=2$ is only an option in Strategy A whereas $A_2=1$ (Coping-Together + Lay Guidance) is shared by both strategies. As Design II has the strictest responder status definition, it leads to the smallest $\hat{p}_1$ out of all competing designs and thus more synthetic participants are deemed non-responders and re-randomized at stage 2. This leads to a larger proportion of individuals assigned $A_2=2$. Thus, the observed effect size in the generated data under Design II tended, on average, to be slightly higher than that of the generated data under the other two designs (despite defining $\delta=0.2$ in Equation~\ref{eq:deltabetasrelationship}). Therefore, the necessary sample size was smallest under Design II. Of course, these results are specific to the example SMART and chosen EDTR comparison. However, this example demonstrates how the proposed simulation framework can be harnessed in practice to compare different designs when a trialist is equipped with data from an initial pilot study.

The sample size estimates for the latter investigation comparing SMARTsize to our simulation procedure, per design, for two different levels of $\delta$ (0.2, 0.5), are shown in Table~\ref{tbl:samplesizes} and Figure~\ref{fig:powercurvealldesigns} (the power estimates per sample size from the simulations are tabulated in the supplementary materials). The simulation results are closer to the sample size calculator results when $\delta=0.5$ rather than $\delta=0.2$, perhaps due to the fact that there is less noise in the simulations at a higher level of $\delta$. For both values of $\delta$, there is no clear pattern with regards to a systematic over- or underestimation of the required sample size. From this, we see that the larger the effect size is likely to be, the less important it may be to capture the uncertainty between the ``fixed" and observed $\delta$. However, if the effect size is small, then it may be more useful to employ a method that reflects this discrepancy in order to better ensure that the required sample size is sufficient, which is especially important in a context where a small effect size is more difficult to detect and thus more likely to lead to a null result simply due to an underpowered design. Hence, in a scenario where $\delta$ is expected to be small, it may be advantageous to employ this simulation-based approach, which incorporates the uncertainty of the observed $\delta$, rather than relying on simpler, and more rigid, sample size formulae. Of course, the likely value of $\delta$ is a judgment call the analyst must make, but this decision can incorporate expert knowledge or previous study results. In many cases, the simulation results here actually suggest a smaller sample size than SMARTsize at $\delta=0.2$, which means the simulation approach does not necessarily always suggest a higher sample size than the calculator and thus does not necessarily suggest a design that would make the trial more costly.

It is also important to note that SMARTsize does not account for potential sharing of treatment options between EDTRs. Effectively, this means that each of the four EDTR comparisons will result in the same sample size calculation, keeping $\delta$ fixed. However, our pedagogical example demonstrates that the observed $\delta$ in the synthetic data is driven by higher proportions of participants re-randomized to a stage 2 treatment option that is not shared by both strategies. Other EDTR comparisons include only treatments that are shared at stage 2 (e.g., Strategy B v.~Strategy C) or only treatments that are not shared at stage 2 (e.g., Strategy A v.~Strategy D). This will lead to different sample size requirements depending on the selected EDTR comparison because the \textbf{observed} $\delta$ will change. The sample size calculator cannot account for this, since it requires the analyst to assume and fix $\delta$.

\begin{table}[H]
\centering
\caption{Estimated sample sizes for Designs I, II, and III using SMARTsize (calculator) and the proposed simulation framework at two power levels (0.8, 0.9) and two effect sizes ($\delta = 0.2, 0.5$). The estimates from the simulation procedure correspond to Figure~\ref{fig:powercurvealldesigns}, where a different data generating mechanism was used for each design (i.e., $\beta_4$ was recalculated to achieve the desired $\delta$ for each design).}
\label{tbl:samplesizes}
\begin{tabular}{clcccc}
\toprule
 &  & \multicolumn{2}{c}{$\mathbf{\delta = 0.2}$} & \multicolumn{2}{c}{$\mathbf{\delta = 0.5}$} \\
\cmidrule(lr){3-4} \cmidrule(lr){5-6}
\textbf{Design} & \textbf{Method} & \textbf{Power = 0.8} & \textbf{Power = 0.9} & \textbf{Power = 0.8} & \textbf{Power = 0.9} \\
\midrule
Design I & Calculator & 1232 & 1648 & 201 & 267 \\
         & Simulation & $(1300, 1400)$ & $(1800, 1900)$ & $(300, 400)$ & $(400, 500)$ \\
\midrule
Design II & Calculator & 1267 & 1694 & 206 & 275 \\
          & Simulation & $(800, 900)$ & $(1100, 1200)$ & $(200, 300)$ & $(200, 300)$ \\
\midrule
Design III & Calculator & 1573 & 2105 & 256 & 341 \\
           & Simulation & $(1100, 1200)$ & $(1500, 1600)$ & $(200, 300)$ & $(400, 500)$ \\
\bottomrule
\end{tabular}
\end{table}

\begin{figure}[H]
    \centering
    \includegraphics[width=\linewidth]{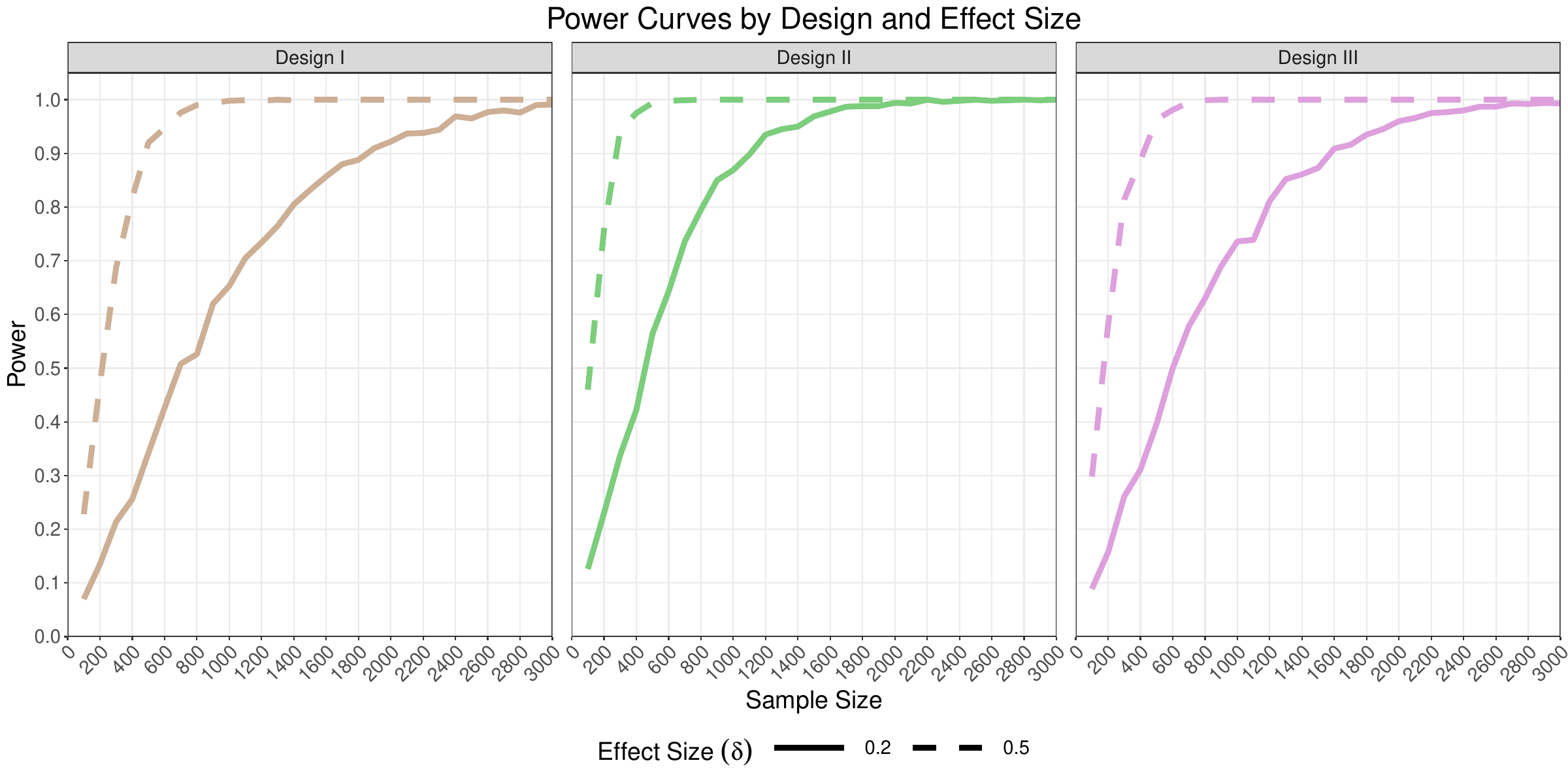}
    \caption{Power curves for Design I (the original pilot design) as well as the two competing designs (Design II in green, Design III in purple), for different effect sizes ($\delta=0.2$ depicted by a solid line, $\delta = 0.5$ depicted by a dashed line).}
    \label{fig:powercurvealldesigns}
\end{figure}

\section{Discussion and Conclusion}
\label{sec:discussion}

We have proposed a novel simulation-based procedure to power designs for a full-scale SMART, comparing EDTRs that differ in stage 1 treatment. The synthetic data generation framework leveraged in this procedure incorporates multiple elements that lend realism and complexity to the synthetic data, including time-varying treatments and covariates as well as missing data. We have also demonstrated two different power analysis investigations: one that fixes the data generating mechanism and estimates potential gains in effect size (and subsequent effects on the minimum required sample size), and another that fixes the effect size and varies the data generating mechanism. Both present great utility to trialists, as this methodology hinges on having already run a pilot trial, in which case it is a natural next step to determine whether the same design should be implemented in a future full-scale SMART (and if so, what is the required sample size), or if it would be advantageous (i.e., would there be a gain in effect size and thus a decrease in the required sample size) in implementing an alternate design.

This simulation tool to power SMARTs addresses a gap in the literature, where there was previously no simulation-based method to power SMARTs to address aim 3. This procedure demonstrates several strengths, most notably that it is flexible and customizable. The sequential nature of generating the synthetic data that one would observe under a given competing design gives the analyst the freedom to model each variable collected in the trial. While we implemented regression at each sequential step, the framework does not limit the potential model choice; one could instead implement, say, random forests, neural networks, or other models provided the available data are large enough. Though our demonstrations involved a continuous outcome, one could easily substitute a binary or time-to-event outcome by re-defining the outcome generation model to handle these distributions. Additionally, the framework is flexible in that there is no limit on the number of stages in the SMART. There are also other design operating characteristics that could be varied within this framework to test competing designs other than responder status. Inclusion and exclusion criteria could be redefined through the assignment of univariate distributions to the R-vine copula fitting. Other trial end-points could be implemented, such as comparing sample size requirements for a design with a continuous outcome versus a dichotomized version. Moreover, we have proposed a way to power SMARTs that does not impose the unrealistic assumptions that $A_1$ subgroups are of equal size and that the response rate is the same. This latter assumption is required not only for the sample size calculators but also other simulation-based methods for powering pilot SMARTs, and as it has been noted that it would be beneficial to develop a method that does not require $p=p_0=p_1$ \citep{Kim2016}, this demonstrates an advantage of the framework proposed in this work. Another simplifying assumption that the sample size formulae require is that the variance of the mean of the responder and non-responder outcomes for a given strategy is greater than the variance of mean outcome among either responders or non-responders. Note that this assumption is satisfied if there is either sufficient separation in the means of the two subgroups (responders v.~non-responders) or no separation but equal variance in the two subgroups (thus meaning that the variance of the overall mean would be the same as the subgroup variances). This is a non-trivial assumption; there could be instances where this is violated if the treatment comparisons involve different modes of treatment (e.g., pharmaceuticals v.~emotional therapy). One can imagine a scenario where both have similar mean outcomes and thus lack distinction between the two groups but one subgroup has much more variation than the other. This would give a scenario where the overall variance would be smaller than the group with more variation (since the overall variance would be essentially an average of the two, with one of them having very small variance).

The investigations in this work have practical applications. For instance, a trialist could perform the first investigation, keeping the data generating mechanism fixed under the original pilot design, and given Figure~\ref{fig:powercurvedeltadist} as an example, there would be evidence to suggest that perhaps redefining the design would be advantageous in terms of minimizing the required sample size to reach a certain level of power. Also, once a design has been selected, a more precise sample size estimate can be derived from these simulations by iteratively updating the range of $n$, identifying smaller and smaller ranges that also become finer and finer. Additionally, in removing the need for the simplifying assumptions of the sample size calculator (which we argue are stronger assumptions than the ones required in the proposed methodology), the simulation-based approach is able to account for uncertainty in $\delta$, thus leading to perhaps more realistic reflections of the necessary sample size since a trial in practice is likely never going to demonstrate an effect size equal to the fixed calculator inputs.

Despite the demonstrated strengths, this work also has some limitations. The sample size of the augmented data set was much larger than that of the internal pilot data only, where the former was used to fit the generative model for shared baseline variables, and the latter was used to fit all subsequent sequential generative models (regression). An implication of this is that the generative performance of variables that exist in only the internal SMART exhibited larger standard errors in the estimated regression parameters, especially since these models are fit to data from a pilot trial, which are small by design. However, the sequential procedure incorporating regression is still more successful at generating realistic synthetic data as compared to other, more common methods in simulation-based power analyses where data are simulated by drawing random samples from pre-defined univariate distributions. The limited number of observations available in a pilot trial also further motivates the choice of regression models, rather than more complex generative models, as fitting more complex models (such as machine learning) require much larger data sets.

In addition, due to the small sample size of the pilot trial, modelling the probability of missingness at each trial stage was impossible (as only three participants dropped out at each stage). This meant we were unable to generate synthetic missingness by predicting the probability of each synthetic participant dropping out and thus implementing synthetic attrition in this data-driven way. Instead, what we implemented here is not so different from the usual fixed inflation factor that is often applied after the sample size is calculated. However, our method does allow for some variation in drop-out per stage and across simulation runs, since participant drop-out is simulated as a random draw from a Bernoulli with probability equal to the empirical estimate, per stage, from the pilot trial. Moreover, depending on the number of designs, this procedure can be resource intensive, thus further motivating the choice of estimating power, at least as a starting point, over a coarser range of $n$. Of course, using a sample size calculator has almost no computational burden, though it requires assumptions that are unlikely to hold in reality. Furthermore, this simulation procedure cannot handle missing data at baseline. This somewhat limited the external data observations that could be selected for augmentation at baseline. Extending this framework to incorporate missing data at baseline would be advantageous, as this would allow for less stringent data augmentation procedures at baseline and would also allow for the inclusion of all internal pilot data in modelling the baseline cohort. Note, however, that fitting elaborate models to capture the real data missingness mechanism at each trial stage does not necessarily serve the goal of the power analysis. Trials generally analyze a very small subset of the collected data, and in the simulation-based power calculation, missingness impacts only response rate and subsequently $A_2$ re-randomization as well as the observed final outcome. Hence, there is a parsimony trade-off, where we aim to build realistic synthetic data (which includes accounting for the real data missingness mechanism), however not all variables collected and generated, which may have missing values, are used in the power calculation and thus the additional model complexity to capture missing data mechanisms for all variables post-baseline would be unnecessary.

Based on the presented results, the proposed simulation-based method perhaps has greater utility when one suspects a smaller effect size, as it is more difficult to detect a smaller effect size. Our results are similar to the sample size calculator for a bigger fixed $\delta$, so the calculator may be sufficient in these scenarios where a larger effect size is likely to be observed. As the flexibility of the proposed simulation-based power analysis lends itself to several extensions, possible future work includes adapting this framework to power SMARTs for aim 4 (estimating the optimal EDTR), which would require re-defining the test statistic as the subgroups would no longer be independent. Additionally, it would be straightforward to expand this framework to incorporate the comparison of competing designs that differ based on design operating characteristics other than responder status, such as inclusion and exclusion criteria or the definition of the outcome. As mentioned in Section~\ref{subsec:simulationmethod}, the flexibility in defining each sequential generative model also allows for other, more complex, model definitions when generating variables collected at the same stage (e.g., at $T_1$). Complexity would be gained had $Z_{11}$ and $Z_{12}$ (and perhaps even $R_1$) been jointly modelled, though whether this would lead to a gain in information and better data generative performance remains unclear. In defining these sequential regression models, it may be advantageous to first verify the causal structure to determine whether variables collected at, say, $T_1$, may, indeed, be strong predictors for other variables collected at $T_1$. To rectify sample size limitations from the pilot trial, it would also be advantageous to incorporate external data at stages after baseline, which may also allow for the modelling of missingness at each stage and thus account for participant attrition in a more sophisticated manner. 

There is also an important trade-off to consider when incorporating external data to generate a synthetic baseline population: extrapolation and generalizability. By incorporating external data at baseline such that the synthetic participants are generated with a baseline multivariate distribution similar to that of the augmented baseline data, we assume that the subsequent generative models for variables post-baseline, which are fit only to internal SMART data, can extrapolate well to the larger synthetic cohort. The closer the external data are to the internal data, the less reliant we are on the assumption that post-baseline data generation models can extrapolate, but the less the synthetic baseline cohort represents a generalizable population of interest. Indeed, an analyst can choose \textit{not} to extrapolate to a larger, and perhaps more generalizable, study population, as this can be controlled through the inclusion and exclusion criteria, and if needed, methods from transportability may also be integrated if distributional similarity is desirable. Hence, it would be interesting to investigate other data augmentation approaches to maximize the augmented data sample size, such as a many-to-one matching procedure, as well as other ways to account for different levels of precision between the internal and external data. As we selected one approach (fixed borrowing) in this work, future work could involve a deeper investigation about the assumptions and implications of different data augmentation strategies within the context of a simulation-based power analysis that uses pilot trial data. If the external data source is less likely to represent individuals who would have feasibly been enrolled in the original pilot trial, then a more sophisticated sampling approach of the external data, or perhaps upweighting or creating multiple copies of the internal SMART data when augmenting baseline data could be considered. Additionally, as a strength of synthetic data generation is the ability to generate as many observations as one desires (up to computational limitations), another potential way to generate a synthetic baseline cohort that limits the reliance on the assumption that the generative models can extrapolate well to a broader study population could be to include an additional variable representing the data source (internal v.~external), and then selecting synthetic baseline observations with ``internal" as the generated value for the synthetic data source variable.

SMARTs are an incredibly useful design to formalize precision medicine's goal of developing individual (dynamic) treatment strategies. In providing an accessible, realistic, and flexible simulation framework to power SMARTs for EDTR comparisons and compare competing study designs, this work has the potential to further popularize SMARTs, allowing for their full potential to be realized.

\section*{Author contributions}

CRediT: \textbf{Niki Z. Petrakos}: Conceptualization, Formal Analysis, Investigation, Methodology, Software, Visualization, Writing -- original draft, Writing -- review \& editing; \textbf{Erica E. M. Moodie} and \textbf{Nicolas Savy}: Conceptualization, Methodology, Supervision, Writing -- review \& editing; \textbf{Manon de Raad}, \textbf{Eric Belzile}, and \textbf{Sylvie Lambert}: Data curation, Resources, Writing -- review \& editing.

\section*{Funding}

Erica E. M. Moodie is a Canada Research Chair (Tier 1) in Statistical Methods for Precision Medicine. This work is supported by a Foundation Grant from the Canadian Institutes of Health Research.

\section*{Supporting information}

Additional details regarding methods and simulation results can be found in the appendix at the end of this article. The pilot SMART data and external RCT data used in this work are not publicly available since they contain sensitive patient information and were obtained under a data-use agreement. Restrictions apply to the availability of these data. Any access to the data requires permission from the collaborating hospital and must comply with applicable ethical, legal, and data-use restrictions. Code for the simulations presented in this paper is available at \url{https://github.com/nikipetrakos/OptimalStudyDesign}.

\bibliographystyle{apalike}  
\bibliography{references}  






\newpage

\appendix

\section*{Appendix}

\section{Data and Model Definitions}

\subsection{Variables}

As noted in the main text, responder definition as used in our investigations was based on a single participant. However, the dyadic definition used in the original pilot trial was the following: if both patient and caregiver had baseline DT$\geq5$, then they were a responder if DT decreased by at least 1 point at stage 1; if only one of the two members had baseline DT$\geq5$, then to be a responder their DT score needed to have decreased by at least one point and the other member's DT score could not have increased by either one point if their baseline score was $\geq5$ or two points if their baseline score was $<5$.

\begin{xltabular}{\textwidth}{l Y}
\caption{Description of variables used to simulate the cancer SMART designs, incorporating both internal SMART data \citep{Lambert2025} and external RCT data \citep{Lambert2021}. }
\label{tab:vardescription} \\

\toprule
\textbf{Variable} & \textbf{Definition and support} \\
\midrule
\endfirsthead

\toprule
\textbf{Variable} & \textbf{Definition and support} \\
\midrule
\endhead

\bottomrule
\endfoot

Participant ID 
& Unique participant identifier (numeric). \\

Age 
& Age in years; continuous, $>45$. \\

Sex 
& Binary (female, male). \\

Marital status 
& Binary (single, not single). \\

Preferred language 
& Binary (English, French). \\

Country of birth 
& Binary (Canada, other). \\

Education 
& Highest level attained; categorical (high school or less, post-secondary, university). \\

DT score (baseline) 
& Distress Thermometer score at baseline; integer (0--10). \\

HADS score (baseline) 
& HADS anxiety score at baseline; continuous (0--21). \\

Treatment (Stage 1) 
& Binary: Coping-Together + Lay Guidance vs.\ Coping-Together only. \\

DT score ($T_1$) 
& Distress Thermometer score at $T_1$ (6 weeks post Stage 1); integer (0--10). \\

HADS score ($T_1$) 
& HADS anxiety score at $T_1$ (6 weeks post Stage 1); continuous (0--21). \\

Responder status ($T_1$) 
& Binary: responder vs.\ non-responder to Stage 1 treatment. \\

Treatment (Stage 2) 
& Categorical: remain on Stage 1 treatment; step up to Coping-Together + Motivational Interviewing; or step up to Coping-Together + Lay Guidance. \\

HADS score ($T_2$) 
& HADS anxiety score at $T_2$ (6 weeks post Stage 2); primary outcome; continuous (0--21). \\

Lost to follow-up 
& Categorical: not lost to follow-up, dropout at $T_1$, or dropout at $T_2$. \\

\end{xltabular}

\subsection{Sequential Generative Models}

Below are the definitions of the models that are fit to sequentially generate the post-randomization variables, as well as baseline covariates not included in the external data, which in our example came from the CanDirect RCT \citep{Lambert2021}. The vector of shared baseline covariates across the internal and external data, $\textbf{X}_{-1}$, includes age, sex, marital status, language, country of birth, and education. The deterministic function used to define responder status at $T_1$ under each of the considered designs is the following:
\begin{itemize}
    \item Design I (the original pilot design): if DT at baseline was $\geq5$ and DT at $T_1$ decreased by at least 1 point, then the participant is a responder. If DT at baseline was $< 5$ and DT at $T_1$ either decreased, stayed the same, or increased by no more than 1 point, then the participant is a responder. Otherwise, the participant is a non-responder.
    \item Design II: if DT and HADS decreased by at least 2 points, then the participant is a responder. Otherwise, the participant is a non-responder.
    \item Design III: if stage 1 DT was less than the $A_1$ subgroup median DT, then the participant is a responder. Otherwise, the participant is a non-responder.
\end{itemize}
For stage 2 treatment, if a participant is a responder, then they remain on stage 1 treatment. If they are a non-responder, then they are re-randomized with equal probability to either a stepped-up care option or the same assigned treatment from stage 1. For example, if a participant was assigned to Coping-Together + Lay-Guidance at Stage 1 and was a non-responder, then they would be re-randomized to either Coping-Together + Motivational Interviewing (stepped-up) or Coping-Together + Lay-Guidance (remain on the same treatment). If a participant was assigned to Coping-Together only at stage 1 and was a non-responder, then they would be re-randomized to either Coping-Together + Lay Guidance (stepped-up) or Coping-Together only (remain on the same treatment). 

\begin{table}[H]
\centering
\caption{Description of data generation models}
\label{tab:execution_models}

\begin{tabularx}{\textwidth}{l l Y}
\toprule
\textbf{Variable to Generate} & \textbf{Model} & \textbf{Covariates} \\
\midrule

Shared variables, baseline ($\textbf{X}_{-1}$) 
& R-vine copula 
& $\mathbf{X}_{-1}$ \\

DT score, baseline ($X_{1}$) 
& Linear regression 
& $\mathbf{X}_{-1}$ \\

Treatment, stage 1 ($A_1$) 
& Stratified, block-randomized 
& HADS $<8$, $8$--$10$, $11$--$21$ \\ 

DT score, $T_1$ ($Z_{11}$) 
& Linear regression 
& $\mathbf{X}, A_1$ \\ 

HADS score, $T_1$ ($Z_{12}$) 
& Linear regression 
& $\mathbf{X}, A_1$ \\ 

Design I responder status, $T_1$ ($R^I_{1}$) 
& Deterministic function 
& $X_{1}, Z_{11}$ \\ 

Design II responder status, $T_1$ ($R^{II}_{1}$) 
& Deterministic function 
& Baseline HADS, $X_{1}, Z_{11}, Z_{12}$ \\ 

Design III responder status, $T_1$ ($R^{III}_{1}$) 
& Deterministic function 
& $Z_{11}$ \\ 

Design I treatment, stage 2 ($A^I_{2}$) 
& Re-randomization design 
& $A_1, R^I_{1}$ \\ 

Design II treatment, stage 2 ($A^{II}_{2}$) 
& Re-randomization design 
& $A_1, R^{II}_{1}$ \\ 

Design III treatment, stage 2 ($A^{III}_{2}$) 
& Re-randomization design 
& $A_1, R^{III}_{1}$ \\ 

Design I HADS score, $T_2$ (outcome, $Y^{I}$) 
& Linear regression 
& $\mathbf{X}, A_1, Z_{11}, Z_{12}, A^{I}_2$ \\ 

Design II HADS score, $T_2$ (outcome, $Y^{II}$) 
& Linear regression 
& $\mathbf{X}, A_1, Z_{11}, Z_{12}, A^{II}_2$ \\ 

Design III HADS score, $T_2$ (outcome, $Y^{III}$) 
& Linear regression 
& $\mathbf{X}, A_1, Z_{11}, Z_{12}, A^{III}_2$ \\ 

\bottomrule
\end{tabularx}
\end{table}

\subsection{Data-Generating Mechanism for Effect Size: Remaining Three Comparisons}

Here we demonstrate the derivations for implementing the data generative effect size for the remaining three strategy comparisons: Strategy A v.~Strategy D, Strategy B v.~Strategy C, and Strategy B v.~Strategy D. Recall that in the synthetic data generation framework, the outcome $Y$ is modelled as $$Y=\beta_0+\beta_1\mathbf{X} + \beta_2A_1+\beta_3\mathds{1}(A_2=1)+\beta_4\mathds{1}(A_2=2)+\epsilon \text{ , with } \epsilon \sim N(0,\sigma^2)$$
where, without loss of generality, $\mathbf{X}$ are participant covariates (baseline variables and stage 1 tailoring variables).

\subsubsection{Strategy A v. Strategy D}

Strategy A imposes $A_1=1$, and $A_2=1$ for responders and $A_2=2$ for non-responders. Strategy D imposes $A_1=0$, and $A_2=0$ for responders and $A_2=0$ for non-responders. Note that no $A_2$ treatment assignment is shared.

Recall the definition of $\delta$ for aim 3: 
\begin{equation}
\label{eq:deltadefrearrangeAvD}
    \delta=\frac{1}{S}\left( \mathbb{E}\left[ Y|A_1=1,A_2=a^{\text{Strat A}}_2 \right] - \mathbb{E}\left[ Y|A_1=0,A_2=a^{\text{Strat D}}_2 \right] \right)
\end{equation}
Similar to the result for Strategy A v. Strategy C as shown in the main paper, the conditional expectation of the outcome for each strategy can be estimated using the outcome model and further conditioning on the response status. Here, the only difference is that $d_2\in\{0\}$ and thus none of $\beta_2,\beta_3,\beta_4$ contribute to the estimation of the conditional expectation of the outcome under Strategy D. 
\begin{equation*}
    \begin{split}
        &\mathbb{E}\left[ Y|A_1=1,A_2=a^{\text{Strat A}}_2 \right]=\beta_0+\beta_1\mathbb{E}[\mathbf{X}]+\beta_2\cdot 1 +\beta_3(p_1)+\beta_4(1-p_1) \\
        &\mathbb{E}\left[ Y|A_1=0,A_2=a^{\text{Strat D}}_2\right]=\beta_0+\beta_1\mathbb{E}[\mathbf{X}]+\beta_2\cdot 0+\beta_3\cdot 0+\beta_4\cdot 0
    \end{split}
\end{equation*}
Subtracting these two conditional expectations and plugging into Equation~\ref{eq:deltadefrearrangeAvD} gives $$\delta\cdot S=\mathbb{E}\left[Y|A_1=1,A_2=a^{\text{Strat A}}_2 \right]-\mathbb{E}\left[ Y|A_1=0,A_2=a^{\text{Strat D}}_2\right]=\beta_2+\beta_3(p_1)+\beta_4(1-p_1).$$
Then fix $\beta_2,\beta_3,\beta_4$ as desired to ensure $\delta$.

\subsubsection{Strategy B v. Strategy C}

Strategy B imposes $A_1=1$, and $A_2=1$ for responders and $A_2=1$ for non-responders. Strategy C imposes $A_1=0$, and $A_2=0$ for responders and $A_2=1$ for non-responders. Note that $A_2=1$ is shared by both strategies.

Recall the definition of $\delta$ for aim 3: 
\begin{equation}
\label{eq:deltadefrearrangeBvC}
    \delta=\frac{1}{S}\left( \mathbb{E}\left[ Y|A_1=1,A_2=a^{\text{Strat B}}_2 \right] - \mathbb{E}\left[ Y|A_1=0,A_2=a^{\text{Strat C}}_2 \right] \right)
\end{equation}
Similar to the result for Strategy A v. Strategy C as shown in the main paper, the conditional expectation of the outcome for each strategy can be estimated using the outcome model and further conditioning on the response status. Here, the only difference is that $b_2\in\{1\}$ and thus $\beta_4$ does not contribute to the estimation of the conditional expectation of the outcome under Strategy B.
\begin{equation*}
    \begin{split}
        &\mathbb{E}\left[ Y|A_1=1,A_2=b_2 \right]=\beta_0+\beta_1\mathbb{E}[\mathbf{X}]+\beta_2\cdot 1 +\beta_3\cdot 1+\beta_4\cdot 0 \\
        &\mathbb{E}\left[ Y|A_1=0,A_2=c_2\right]=\beta_0+\beta_1\mathbb{E}[\mathbf{X}]+\beta_2\cdot 0+\beta_3(1-p_0)+\beta_4\cdot 0
    \end{split}
\end{equation*}
Subtracting these two conditional expectations and plugging into Equation~\ref{eq:deltadefrearrangeBvC} gives $$\delta\cdot S=\mathbb{E}\left[Y|A_1=1,A_2=a^{\text{Strat B}}_2 \right]-\mathbb{E}\left[ Y|A_1=0,A_2=a^{\text{Strat C}}_2\right]=\beta_2+\beta_3(p_0).$$
Then fix $\beta_2,\beta_3$ as desired to ensure $\delta$.

\subsubsection{Strategy B v. Strategy D}

Strategy B imposes $A_1=1$, and $A_2=1$ for responders and $A_2=1$ for non-responders. Strategy D imposes $A_1=0$, and $A_2=0$ for responders and $A_2=0$ for non-responders. Note that no $A_2$ treatment assignment is shared.

Recall the definition of $\delta$ for aim 3: 
\begin{equation}
\label{eq:deltadefrearrangeBvD}
    \delta=\frac{1}{S}\left( \mathbb{E}\left[ Y|A_1=1,A_2=a^{\text{Strat B}}_2 \right] - \mathbb{E}\left[ Y|A_1=0,A_2=a^{\text{Strat D}}_2 \right] \right)
\end{equation}
Similar to the result for Strategy A v. Strategy C as shown in the main paper, the conditional expectation of the outcome for each strategy can be estimated using the outcome model and further conditioning on the response status. Here, $b_2\in\{1\}$ and thus $\beta_4$ does not contribute to the estimation of the conditional expectation of the outcome under Strategy B, and $d_2\in\{0\}$ and thus none of $\beta_2,\beta_3,\beta_4$ contribute to the estimation of the conditional expectation of the outcome under Strategy D.
\begin{equation*}
    \begin{split}
        &\mathbb{E}\left[ Y|A_1=1,A_2=a^{\text{Strat B}}_2 \right]=\beta_0+\beta_1\mathbb{E}[\mathbf{X}]+\beta_2\cdot 1 +\beta_3\cdot 1+\beta_4\cdot 0 \\
        &\mathbb{E}\left[ Y|A_1=0,A_2=a^{\text{Strat D}}_2\right]=\beta_0+\beta_1\mathbb{E}[\mathbf{X}]+\beta_2\cdot 0+\beta_3\cdot 0+\beta_4\cdot 0
    \end{split}
\end{equation*}
Subtracting these two conditional expectations and plugging into Equation~\ref{eq:deltadefrearrangeBvD} gives $$\delta\cdot S=\mathbb{E}\left[Y|A_1=1,A_2=a^{\text{Strat B}}_2 \right]-\mathbb{E}\left[ Y|A_1=0,A_2=a^{\text{Strat D}}_2\right]=\beta_2+\beta_3.$$
Then fix $\beta_2,\beta_3$ as desired to ensure $\delta$.

\section{Additional Figures and Results}

\subsection{Density Plots: Baseline Variables from Synthetic Data, SMART (internal data), and RCT (external data)}

\begin{figure}[H]
    \centering
    \includegraphics[width=0.5\linewidth]{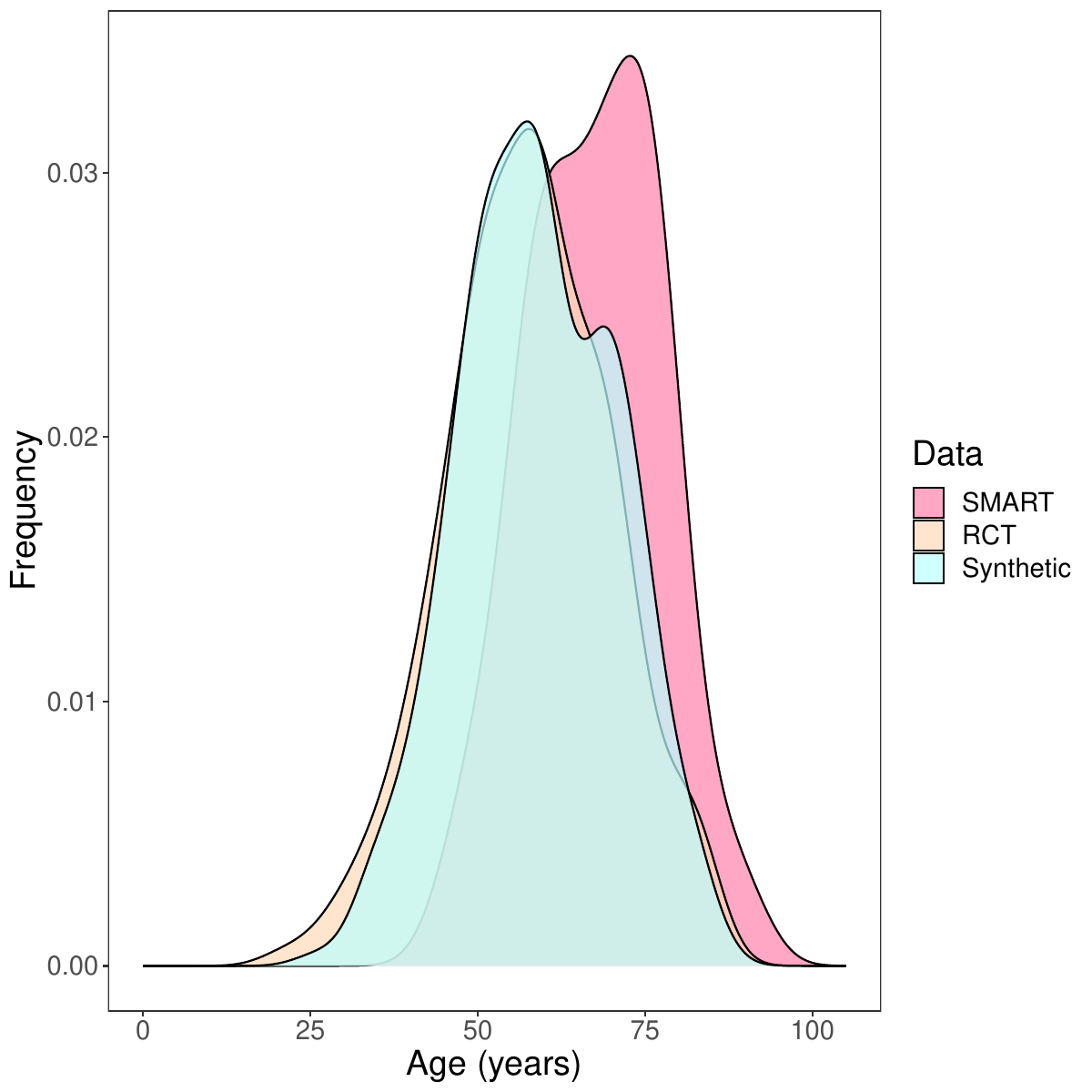}
    \caption{Univariate density plot of age in the baseline Cancer SMART complete cases (pink), the RCT complete cases (peach), and one synthetic data set (blue).}
\end{figure}


\begin{figure}[H]
    \centering
    \includegraphics[width=0.5\linewidth]{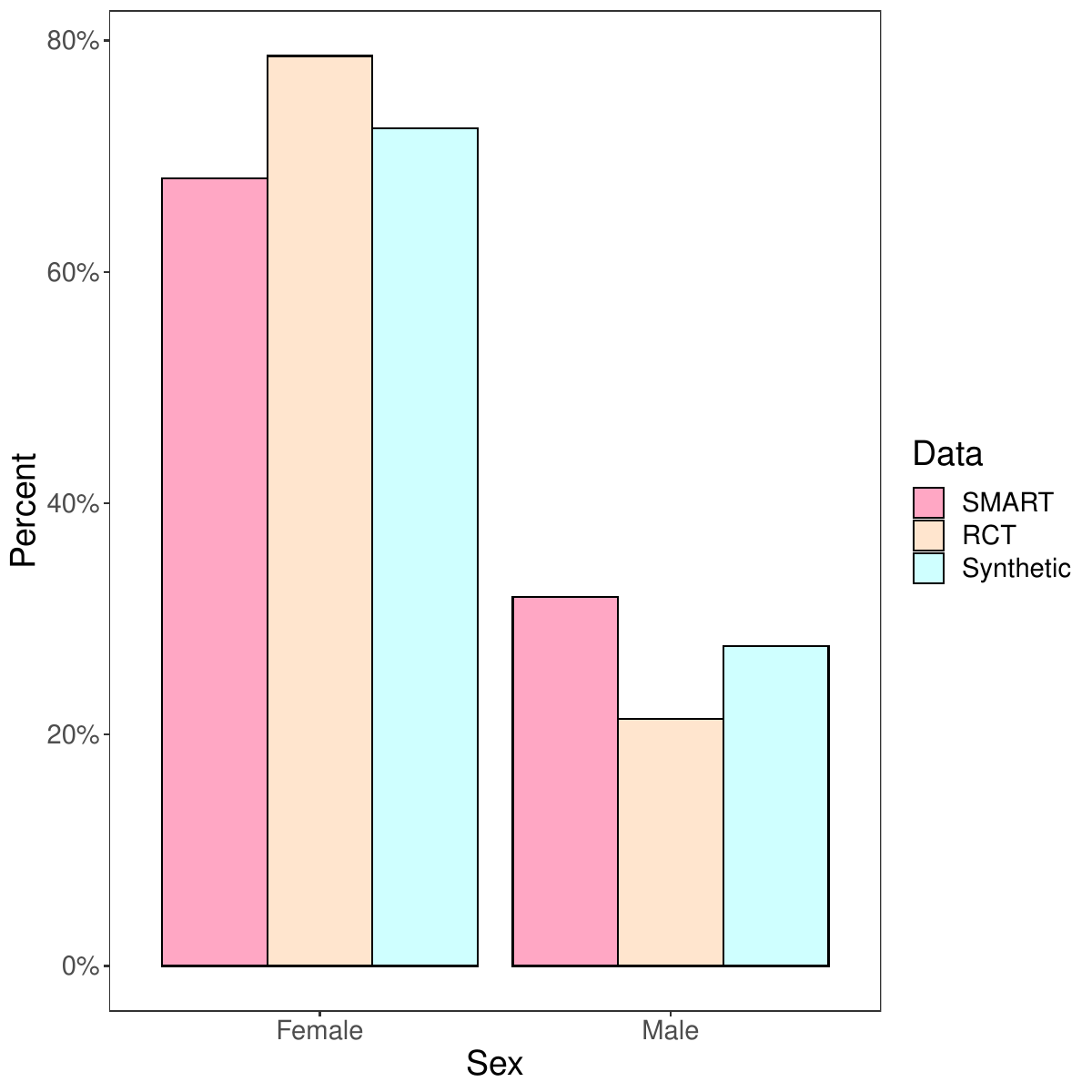}
    \caption{Univariate density plot of sex in the baseline Cancer SMART complete cases (pink), the RCT complete cases (peach), and one synthetic data set (blue).}
\end{figure}

\begin{figure}[H]
    \centering
    \includegraphics[width=0.5\linewidth]{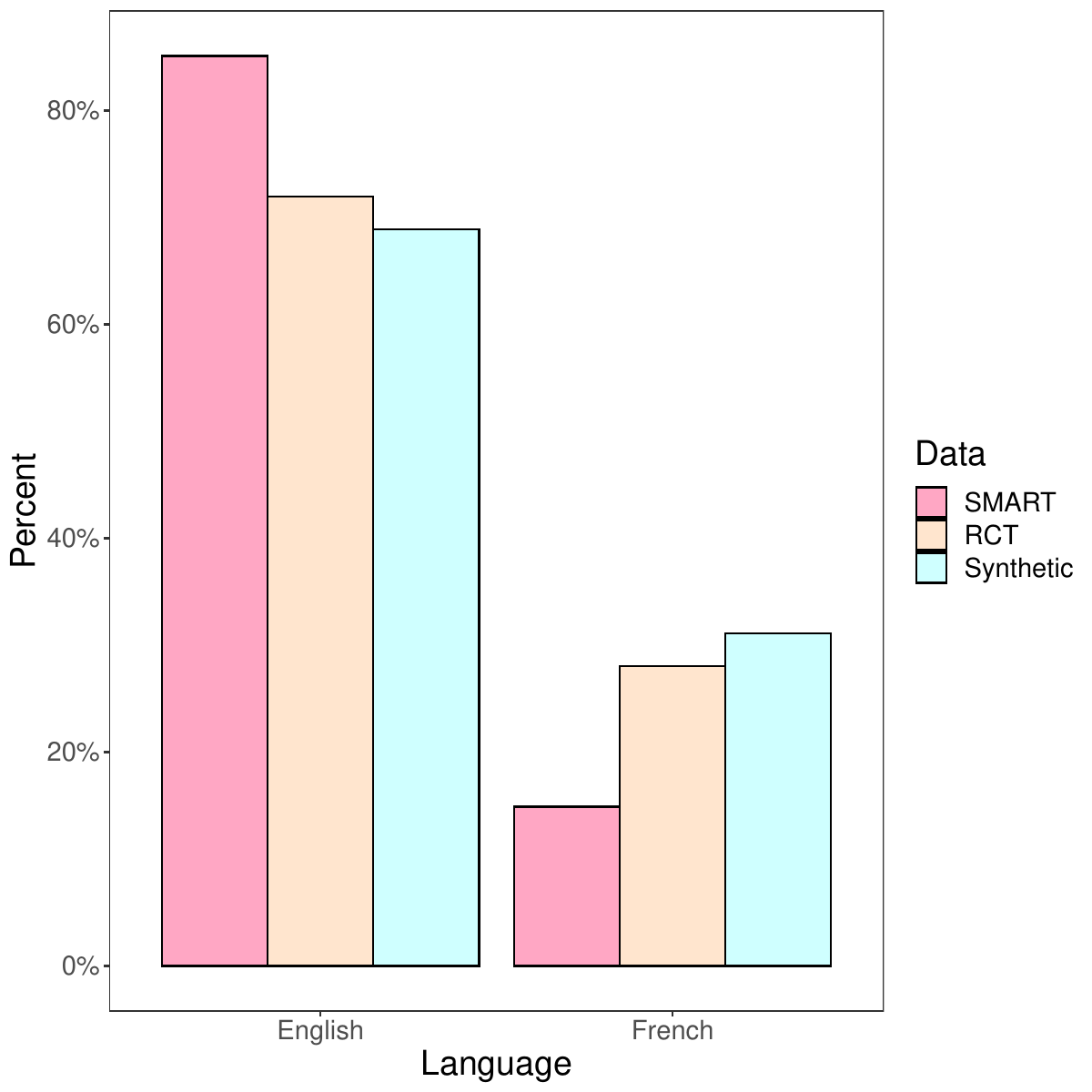}
    \caption{Univariate density plot of language in the baseline Cancer SMART complete cases (pink), the RCT complete cases (peach), and one synthetic data set (blue).}
\end{figure}

\begin{figure}[H]
    \centering
    \includegraphics[width=0.5\linewidth]{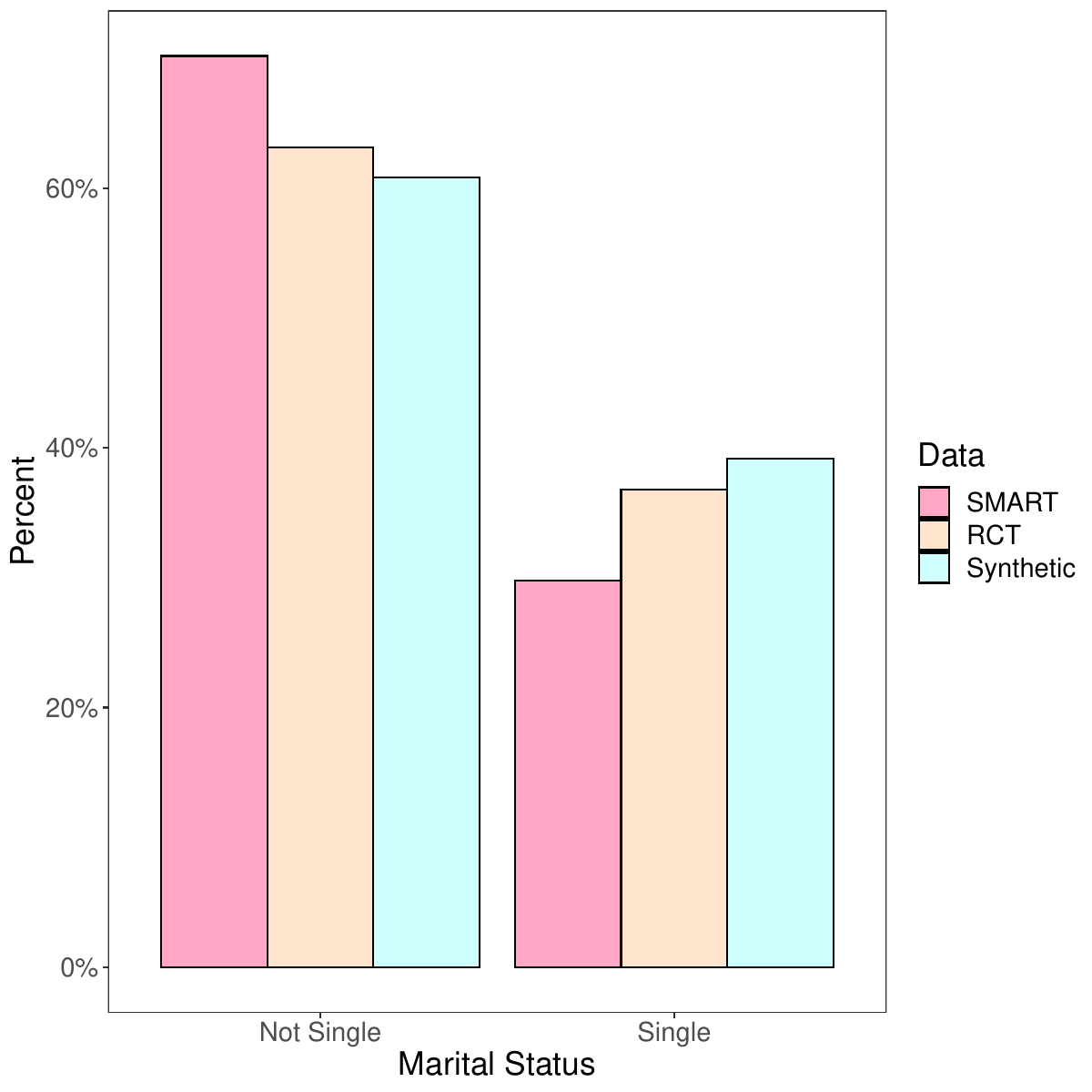}
    \caption{Univariate density plot of marital status in the baseline Cancer SMART complete cases (pink), the RCT complete cases (peach), and one synthetic data set (blue). Data were coarsened so that categories better matched across the internal and external data (i.e., Married and Common Law were merged as Not Single, while Separated, Divorced, Widowed, and Single/Never Married were merged as Single). The Single/Never Married stratum was observed only in the RCT (and not in the SMART). The RCT originally had missing marital status values whereas this variable was completely observed in the SMART.}
\end{figure}

\begin{figure}[H]
    \centering
    \includegraphics[width=0.5\linewidth]{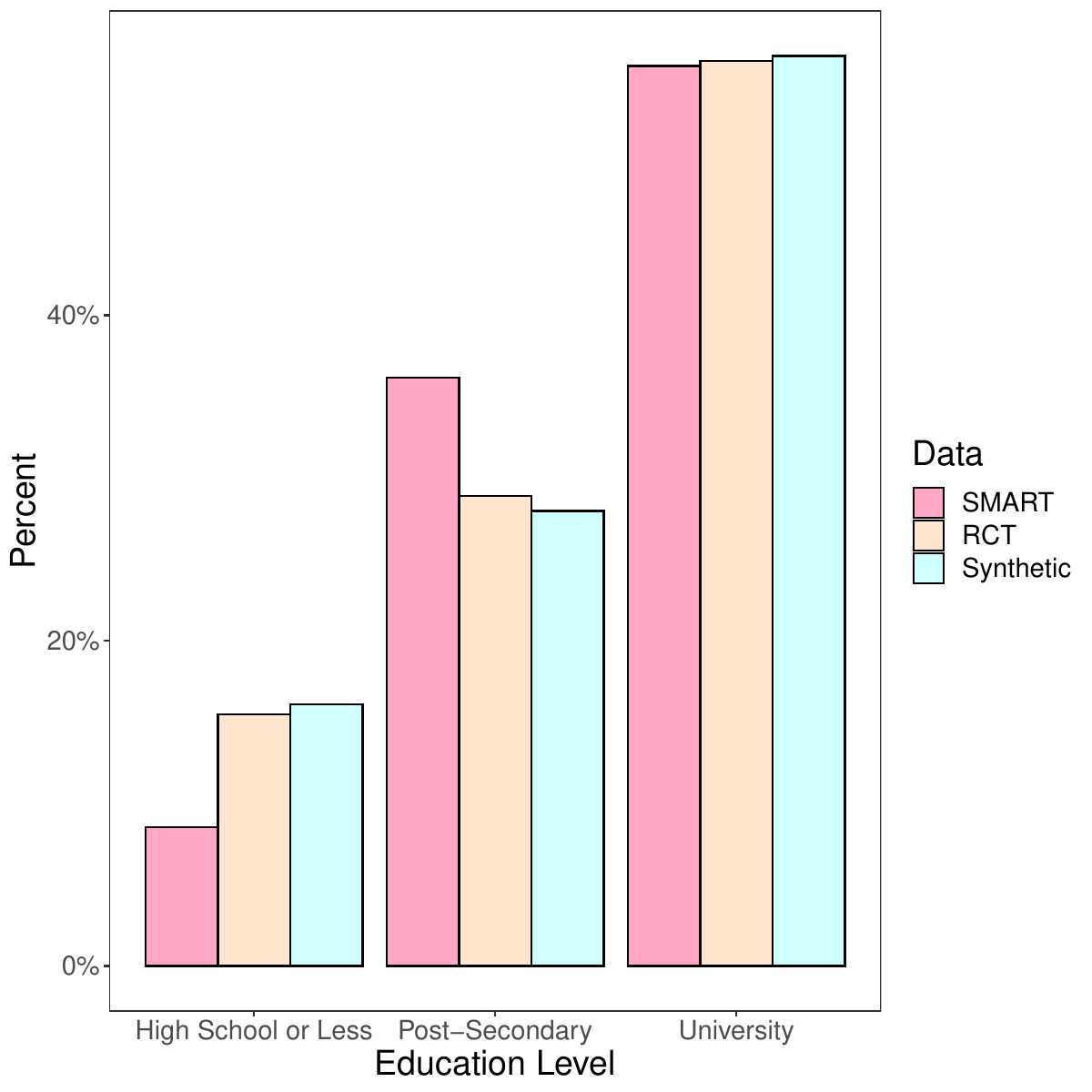}
    \caption{Univariate density plot of highest level of education in the Cancer SMART complete cases (pink), the RCT complete cases (peach), and one synthetic data set (blue). Two of the categories in the SMART data were merged to better match the RCT: Undergraduate University Degree and Graduate Diploma were merged as University. Both the SMART and the RCT originally had missing education values.}
\end{figure}

\begin{figure}[H]
    \centering
    \includegraphics[width=0.5\linewidth]{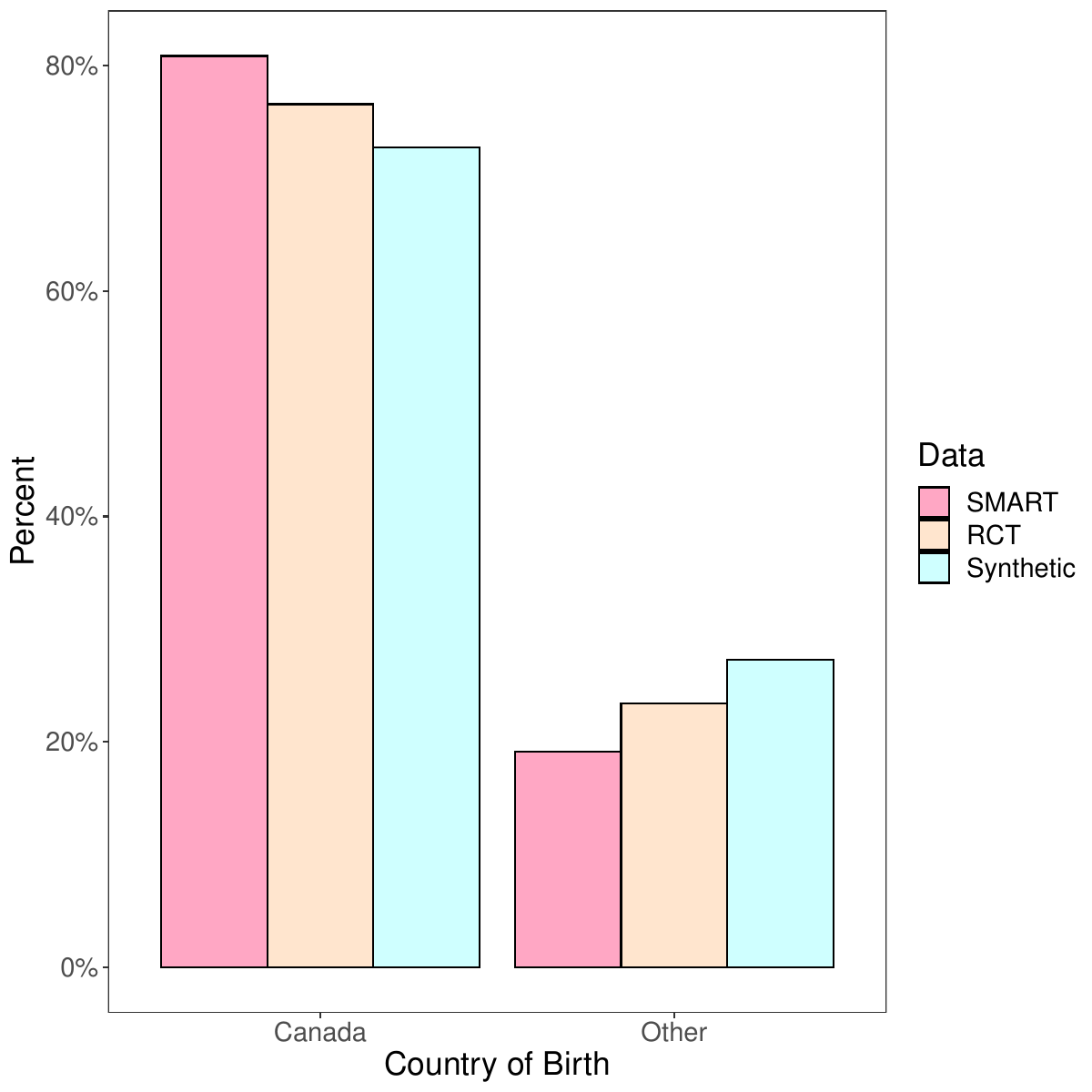}
    \caption{Univariate density plot of country of birth in the Cancer SMART complete cases (pink), the RCT complete cases (peach), and one synthetic data set (blue). The RCT originally had missing country of birth values whereas this variable was completely observed in the SMART.}
\end{figure}

\begin{figure}[H]
    \centering
    \includegraphics[width=0.5\linewidth]{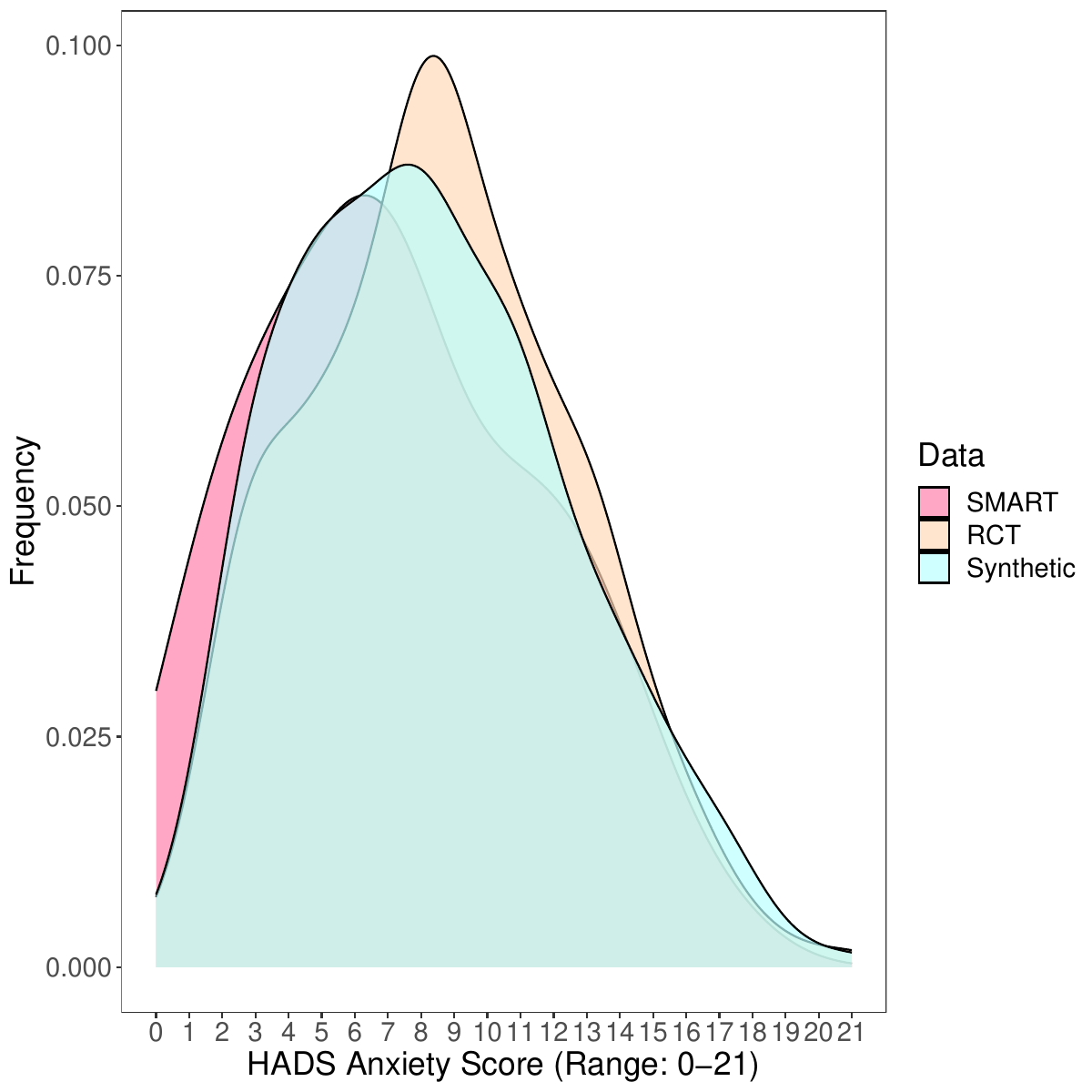}
    \caption{Univariate density plot of (continuous) HADS anxiety score at baseline in the Cancer SMART complete cases (pink), the RCT complete cases (peach), and one synthetic data set (peach). The RCT originally had missing values for this variable blue this variable was completely observed in the SMART.}
\end{figure}

\subsection{Simulation Results}

\begin{table}[H]
\centering
\caption{Estimated power per sample size tested (100, 200, ..., 3000) for the original pilot design with two other competing designs using the proposed simulation framework at two power levels (0.8, 0.9) and two effect sizes ($\delta = 0.2, 0.5$). The estimates from the simulation procedure correspond to the first investigation, where a different data generating mechanism was used for each design (i.e., $\beta_4$ was re-defined per design).}
\label{tbl:samplesizes_wide}
\begin{tabular}{c|cc|cc|cc}
\toprule
& \multicolumn{6}{c}{\textbf{Power}} \\
\cmidrule(lr){2-7}
& \multicolumn{2}{c|}{\textbf{Design I}} 
& \multicolumn{2}{c|}{\textbf{Design II}} 
& \multicolumn{2}{c}{\textbf{Design III}} \\
\textbf{Sample Size} 
& $\delta=0.2$ & $\delta=0.5$ 
& $\delta=0.2$ & $\delta=0.5$ 
& $\delta=0.2$ & $\delta=0.5$ \\
\midrule

100  & 0.070 & 0.228 & 0.126 & 0.460 & 0.089 & 0.298 \\
200  & 0.135 & 0.471 & 0.230 & 0.757 & 0.157 & 0.580 \\
300  & 0.214 & 0.687 & 0.338 & 0.943 & 0.261 & 0.814 \\
400  & 0.256 & 0.817 & 0.422 & 0.975 & 0.311 & 0.887 \\
500  & 0.342 & 0.920 & 0.564 & 0.994 & 0.396 & 0.963 \\
600  & 0.426 & 0.947 & 0.643 & 0.998 & 0.500 & 0.981 \\
700  & 0.508 & 0.976 & 0.736 & 0.999 & 0.578 & 0.995 \\
800  & 0.526 & 0.990 & 0.795 & 1.000 & 0.630 & 0.999 \\
900  & 0.620 & 0.994 & 0.850 & 1.000 & 0.690 & 1.000 \\
1000 & 0.653 & 0.998 & 0.869 & 1.000 & 0.736 & 1.000 \\
1100 & 0.705 & 0.999 & 0.898 & 1.000 & 0.739 & 1.000 \\
1200 & 0.734 & 0.998 & 0.935 & 1.000 & 0.811 & 1.000 \\
1300 & 0.765 & 1.000 & 0.945 & 1.000 & 0.852 & 1.000 \\
1400 & 0.805 & 0.999 & 0.950 & 1.000 & 0.861 & 1.000 \\
1500 & 0.832 & 1.000 & 0.969 & 1.000 & 0.873 & 1.000 \\
1600 & 0.857 & 1.000 & 0.978 & 1.000 & 0.909 & 1.000 \\
1700 & 0.880 & 1.000 & 0.987 & 1.000 & 0.916 & 1.000 \\
1800 & 0.888 & 1.000 & 0.988 & 1.000 & 0.935 & 1.000 \\
1900 & 0.910 & 1.000 & 0.988 & 1.000 & 0.945 & 1.000 \\
2000 & 0.922 & 1.000 & 0.994 & 1.000 & 0.960 & 1.000 \\
2100 & 0.937 & 1.000 & 0.993 & 1.000 & 0.966 & 1.000 \\
2200 & 0.938 & 1.000 & 1.000 & 1.000 & 0.975 & 1.000 \\
2300 & 0.944 & 1.000 & 0.996 & 1.000 & 0.977 & 1.000 \\
2400 & 0.969 & 1.000 & 0.998 & 1.000 & 0.980 & 1.000 \\
2500 & 0.965 & 1.000 & 1.000 & 1.000 & 0.987 & 1.000 \\
2600 & 0.977 & 1.000 & 0.998 & 1.000 & 0.987 & 1.000 \\
2700 & 0.980 & 1.000 & 0.999 & 1.000 & 0.993 & 1.000 \\
2800 & 0.976 & 1.000 & 1.000 & 1.000 & 0.992 & 1.000 \\
2900 & 0.990 & 1.000 & 0.999 & 1.000 & 0.994 & 1.000 \\
3000 & 0.991 & 1.000 & 1.000 & 1.000 & 0.993 & 1.000 \\

\bottomrule
\end{tabular}
\end{table}

\end{document}